\documentclass[pre,superscriptaddress, nofootinbib]{revtex4-2}

\usepackage[utf8]{inputenc} 
\usepackage[T1]{fontenc}    
\usepackage{hyperref}       
\usepackage{url}            
\usepackage{booktabs}       
\usepackage{amssymb,amsmath,amsfonts}       
\usepackage{nicefrac}       
\usepackage{graphicx}       
\graphicspath{{./grafici/}} 
\usepackage{caption,subcaption}
\usepackage{float}
\usepackage{placeins} 
\usepackage{utf8-symbols} 

\usepackage{xcolor}

\begin{document}

\title{The most uniform distribution of points on the sphere}

\author{Luca Maria Del Bono}
\affiliation{Dipartimento di Fisica, Sapienza Università di Roma, Piazzale Aldo Moro 5, Rome 00185, Italy}

\author{Flavio Nicoletti}
\affiliation{Dipartimento di Fisica, Sapienza Università di Roma, Piazzale Aldo Moro 5, Rome 00185, Italy}

\author{Federico Ricci Tersenghi}
\affiliation{Dipartimento di Fisica, Sapienza Università di Roma, Piazzale Aldo Moro 5, Rome 00185, Italy}
\affiliation{CNR-Nanotec, Rome unit and INFN, sezione di Roma1, Piazzale Aldo Moro 5, Rome 00185, Italy}

\begin{abstract}
How to distribute a set of points uniformly on a spherical surface is a very old problem that still lacks a definite answer. In this work, we introduce a physical measure of uniformity based on the distribution of distances between points, as an alternative to commonly adopted measures based on interaction potentials. We then use this new measure of uniformity to characterize several algorithms available in the literature. We also study the effect of optimizing the position of the points through the minimization of different interaction potentials via a gradient descent procedure. In this way, we can classify different algorithms and interaction potentials to find the one that generates the most uniform distribution of points on the sphere.
\end{abstract}

\maketitle

\section{Introduction}

Distributing a finite set of points uniformly on a spherical surface is a fundamental problem, dating back at least the beginning of the XX century \cite{thomson1904xxiv,2b51ac61c9fb43a29f181b14458a1f5e}. It has applications, among others, in biology \cite{barreira2011surface, PhysRevLett.98.185502}, engineering \cite{944811} and astronomy \cite{bauer2000distribution} and is, in general, a relevant question every time a discretization of the sphere is required. However, the first difficulty one encounters when dealing with this kind of problem is how to properly define uniformity.

A first definition of uniform distribution of a set of points on the sphere focuses on the approximation of integrals on the sphere
\cite{beentjes2015quadrature}. Indeed, one can search for a distribution of points such that the difference between the numerical integration carried on using these points and the analytic result is small. Expanding on this idea, one can define $t-$spherical designs as sets of points that allow for perfect integration (i.e.\ the numerical and analytical results coincide) for all polynomials up to degree $t$ \cite{saff1997distributing}.
 
Another very studied and rather classical definition of uniformity for a set of $N_p$ points is based on the minimization of the interaction potential among these points. The first problem of this kind was proposed by Thomson who considered $N_p$ electrons on the sphere interacting through a Coulomb electric potential \cite{thomson1904xxiv}. The Thomson problem has been recently generalized by taking into account the possibility of electrons being inside the sphere \cite{Y.Levin_2003} and interacting through different potentials decaying as $r^{-\alpha}$ at large distance \cite{PhysRevLett.89.185502}. A more geometrical definition of uniformity focuses on packing a given number of circles on the sphere such that the minimum radius is maximized: this is the so-called Tammes problem \cite{2b51ac61c9fb43a29f181b14458a1f5e}, and can be seen as the $\alpha\to\infty$ limit of the Thomson problem.

On the one hand, the use in the Thomson problem of smooth and soft potentials (e.g.\ the Coulomb one or anyone with $\alpha$ not too large) allows for a direct application of standard optimization techniques, such as the Gradient Descent (GD) and the quasi-Newton method. On the other hand, the sphere packing approach in the Tammes problem would be better to generate points uniformly on the sphere, as it does not require the introduction of an arbitrary potential, but its solution has a much higher computational cost. Our analysis suggests a way to bridge the gap between the theoretical optimum and what is computationally achievable.

Given the arbitrariness in the definition of uniformity, over the years many sphere discretization algorithms have been proposed, all developed with different goals in mind. Algorithms for finding the ground state configurations of a given potential range from more geometrical approaches \cite{altschuler1997possible} to a brute-force optimization of many randomly generated configurations \cite{erber1997complex}, passing through more sophisticated procedures such as Simulated Annealing \cite{simAnnealing}. Algorithms developed for practical applications may have additional constraints due to the specific use they are intended for. This is the case of sphere discretizations used in the analysis of astronomical data \cite{Gorski_2005, malkin2019new}. In the present work, we are going to analyze only a small number of such algorithms, but our criteria can be applied to any other procedure.

Our definition of uniformity is based on quantifying how crystalline-like the distribution of points is.
In the Euclidean space, a regular lattice can be straightforwardly defined through discretized transformations, e.g.\ translation by lattice vectors.
However, on a sphere such regular lattices do not exist, but for few values of $N_p$.
So, in general, the most uniform distribution of points on a sphere is not a perfectly regular lattice and contains defects.
This is the essential reason why the definition of the most uniform distribution of points on a sphere is not easy to provide.

Our criteria for uniformity will consider first the minimization of defects and then the distance between points, especially nearest neighbors.
Indeed we aim primarily at controlling the discretization of the sphere \emph{locally}.
This is a crucial requirement when one is interested in approximating a function defined on the sphere with function values computed on a discrete set of points.
An illuminating application requiring such a uniform local discretization is the approximation of marginals for Heisenberg models that we have solved recently \cite{del2024compute}.

As expected, if one is interested in optimizing the nearest neighbors' distances, short-ranged potentials (i.e.\ large $\alpha$ values) are better and provide locally improved crystalline order. Under this new uniformity criterion, long-ranged potentials (e.g.\ the Coulomb one) are very inefficient, because any given point interacts too much with far away points and this is not useful at all to optimize the discretization locally. Thanks to our analysis, we can provide a recipe to both generate distributions at the state-of-the-art level and to benchmark future methods, thus allowing us to study the most uniform distribution of points on the sphere.

It is important to stress that spatially uniform distributions are different from the so-called hyperuniform distributions. Indeed, the concept of hyperuniformity on the sphere, which will be described in detail in Sec.~\ref{sec:hyperuniformity}, is a long-distance property of the system, which does not take much into account local properties of the systems, as proven by the huge variety of hyperuniform points distributions found in previous works \cite{bovzivc2019spherical, meyra2019hyperuniformity}. Our results further confirm this claim.

This paper is organized as follows. In Sec.~\ref{sec:background} we summarise known results and give some definitions that will be used throughout the rest of the manuscript. In particular, we define a defect in a distribution of points on the sphere (Sec.~\ref{sec:defects}), we define the different measures of uniformity one can introduce (Sec.~\ref{sec:meas_unif}) as well as the concept of hyperuniformity (Sec.~\ref{sec:hyperuniformity}). In Sec.~\ref{sec:methods} we introduce the methods we use in this work. Specifically, we briefly describe the algorithms analyzed in this paper (Sec.~\ref{sec:algorithms}) and some possible choices of the potentials for applying a gradient descent optimization procedure (Sec.~\ref{sec:potentials}). In Sec.~\ref{sec:results} we report the original results we have obtained and point out what is the best method to distribute points uniformly on the sphere. Specifically, we compare the different algorithms to a random distribution (Sec.~\ref{sec:comparison_random}), we compute the distribution of distances (Sec.~\ref{sec:dist_distr}), we check whether the methods satisfy hyperuniformity criteria (Sec.~\ref{sec:hyper_results}), we study thoroughly the first peak of the distribution of distances to find the most uniform grid (Sec.~\ref{sec:studyfirstpeak}), we perform extrapolation to infinitely sharp potentials and to an infinite number of points (Sec.~\ref{sec:extrapolation}) and we check stability against perturbations (Sec.~\ref{sec:stability}). Finally, in Sec.~\ref{sec:conclusions} we draw our conclusion and summarize the main results of the paper.

\section{Theoretical Background}\label{sec:background}

\subsection{Defects}\label{sec:defects}

It is well-known that in the plane the highest density packing of circles is the hexagonal one \cite{chang2010simple}. A perfectly hexagonal lattice cannot, however, be embedded on the sphere surface. Indeed \cite{saff1997distributing}, from Euler's formula
\begin{equation}
    \mathcal{F} + \mathcal{V} =  \mathcal{E} + 2,
\end{equation}
($\mathcal{F}$, $\mathcal{V}$ and $\mathcal{E}$ being the number of faces, vertices and edges, respectively) and assuming that faces are triangles whose vertices only have 4, 5, 6 or 7 neighbors, it follows that
\begin{equation}\label{eq:defects}
    2\mathcal{V}_4 + \mathcal{V}_5 - \mathcal{V}_7 = 12,  
\end{equation}
where $\mathcal{V}_i$ is the number of vertices with $i$ neighbours. Here, the neighborhood can be defined in terms of the Voronoi diagrams. Given a set of points $\{\mathcal{P}_i\}$ on the sphere, the Voronoi decomposition is obtained associating to each point the region $\mathcal{R}_i$ of all the points on the sphere which are closer to $\mathcal{P}_i$ than to any other $\mathcal{P}_{j \neq i}$.
As a consequence, it is impossible not to have any defects (points with more or less than six neighbors, also called \textit{disclinations} \cite{wales2006structure}). Moreover, from \eqref{eq:defects} follows that, if $\mathcal{V}_4 = \mathcal{V}_7 = 0$ then $\mathcal{V}_5 = 12$. The most symmetric property that we can require from these twelve minimal defects is that they are placed at the vertices of a regular icosahedron. Configurations that have this property are called \emph{icosadeltahedral}. Algorithms such as the Lattice Points (LP) method (see Sec.~\ref{sec:algorithms}) generate configuration with this minimal number of defects, while the Polar Coordinates Subdivision method does not.

It has been shown that icosadeltahedral configuration, and in particular configurations generated through LP, are not generally the lowest energy configuration, e.g.\ for the Coulomb potential \cite{perez1997influence, katanforoush2003distributing}, despite the low-energy configurations still displaying symmetries when defects are taken into account \cite{perez1999symmetric}. Indeed, optimal configurations start to develop more defects the more points are placed on the sphere, usually near each other in the form of \textit{dislocations} \cite{wales2009defect}, and for large values of $N_p$ only a very small fraction of configurations is defect-free \cite{altschuler2006defect}.

Since grids that minimize Coulomb energy usually have many defects, and since we expect that a uniform mesh should have a structure that is locally as close as possible to that of a hexagonal lattice, it follows that minimizing the Coulomb energy might not be the best descriptor of uniformity for practical purposes. One possible reason could be that the Coulomb interaction is long-ranged and, thus, allows points that are very far away to interact with each other. On the other hand, in a hexagonal lattice points should mainly interact with neighbors. We shall later show evidence in favor of this intuition by comparing different potentials.

\subsection{Measures of uniformity}\label{sec:meas_unif}
As already mentioned, there is no unique way to define how uniform a distribution of points on the sphere is. One common approach is to define a potential and then try to find the distribution that minimizes (or, in some special cases, maximizes) it.
Common choices for the energy \cite{alishahi2015spherical} are the aforementioned Coulomb potential\footnote{Sometimes, alternative definitions with sums running over $i < j$ are used. The two choices are obviously equivalent, apart from an overall $1/2$ factor.}
\begin{equation}
    E_\text{Coul} \equiv \sum_{i<j} \frac{1}{|\Vec{r}_i - \Vec{r}_j|},
\end{equation}
where the sum runs over all couples of different points and the distance is the Euclidean distance; its generalization (Riesz $s-$energy)
\begin{equation}
    E_\text{s} \equiv \sum_{i<j} \frac{1}{|\Vec{r}_i - \Vec{r}_j|^s};
\end{equation}
and the minimum spacing (related to the Tammes problem)
\begin{equation}
    E_\text{ms} \equiv \min_{i<j} |\Vec{r}_i - \Vec{r}_j|.
\end{equation}
Other frequently used potentials are the logarithmic and harmonic energies:
\begin{equation}
    E_\text{log} \equiv \sum_{i<j} \log \Big [ \frac{1}{|\Vec{r}_i - \Vec{r}_j|} \Big ],
\end{equation}
\begin{equation}
    E_\text{har} \equiv \sum_{i<j} |\Vec{r}_i - \Vec{r}_j|^2.
\end{equation}
The problem then becomes that of finding the set of points that minimizes (in the minimum spacing and harmonic cases, maximizes) the energy. One of the advantages of these procedures is that there are results (such as the \textit{Poppy-seed Bagel} theorem) that guarantee that minimal energy configurations of suitably chosen potentials are well-behaved in the asymptotic limit.

In this work, we have instead chosen to focus mainly on the distribution $g(r)$ of distances $r$ between pairs of different points,
\begin{equation}
    g(r) \equiv \frac{1}{N_p(N_p-1)} \sum_{i, j} \delta\left(r-|\vec{r}_i -\Vec{r}_j|\right).
\end{equation}
From a practical point of view, $g(r)$ is more commonly studied as a histogram, so that the $\delta$ centered at close-by distances are coarse-grained together, resulting in a smoothed-out behavior of the function.
The use of $g(r)$ is connected to the theory of crystallography, in which the structure factor is connected to the charge distribution of ions in the crystal \cite{Ashcroft76}. Indeed, in a perfect crystal, despite the coarse-graining,  $g(r)$ is still the sum of many $\delta$ functions centered at the distances between first neighbors, second neighbors, third neighbors, and so on. On the sphere, on the other hand, the peaks are widened due to the curvature and the presence of defects. Despite this, we expect that the more the distribution of points is regular, i.e.\ the more it locally resembles a crystal, the narrower the peaks will be. Since a crystalline-like configuration would be the most uniform, we expect that the better-behaving distributions are those for which $g(r)$ is closer to that of a crystal.

To quantify this approach, we have studied the first peak of the $g(r)$, i.e.\ the one connected to distances between first neighbors. Subsequent peaks can be studied, but the analysis becomes noisier due to the increasing number of neighbors (e.g.\ the second peak includes both points at distance 2 and at distance $\sqrt{3}$, in units of first neighbors distance) and for large $r$ it becomes altogether impossible to separate different peaks.


\subsection{Hyperuniformity}\label{sec:hyperuniformity}

The concept of hyperuniformity was first introduced approximately twenty years ago in the Euclidean space setup. It has recently been applied to distributions of points on the sphere. 

In Euclidean space, hyperuniformity is identified by a vanishing structure factor $S$ at low wave vectors $\Vec{k}$, i.e. by
\begin{equation}\label{eq:hyperuniform}
    \lim_{|\Vec{k}| \to 0}S(\Vec{k}) = 0.
\end{equation}
On the sphere, the structure factor of a distribution of $N_p$ points is \cite{bovzivc2019spherical}:
\begin{equation}\label{eq:s_factor}
    S(\ell) = \frac{1}{N_p}\sum_{i, j = 1}^{N_p} P_\ell(\cos\gamma_{ij}),
\end{equation}
$\gamma_{ij}$ being the spherical distance between points $i$ and $j$. $P_\ell$ are the Legendre polynomials. In this case, the wave vector $\Vec{k}$ is substituted by the wave number $\ell$ and therefore the condition in \eqref{eq:hyperuniform} becomes the requirement that the structure factor vanishes at low $\ell$. It is important to notice that, while $\Vec{k}$ is a continuous variable, $\ell$ is discrete-valued and therefore the limit $\ell \to 0$ is not well-defined, so the uniformity criterion used in practice is that the structure factor in \eqref{eq:s_factor} becomes zero for sufficiently small $\ell$.

An additional definition of hyperuniformity on the sphere can be given in terms of the \textit{spherical cap variance}, $\sigma_N^2(\theta)$. For a given angle $\theta$, the spherical cap variance is defined as 
\begin{equation}
    \sigma_N^2(\theta) \equiv \langle N(\theta, \Vec{r})^2 \rangle_{\Vec{r} \in \mathbb{S}^2} - \langle N(\theta, \Vec{r}) \rangle_{\Vec{r} \in \mathbb{S}^2}^2,
\end{equation}
where $N(\theta, \Vec{r})$ is the number of points in the spherical cap of opening angle $\theta$ and centered in the point $\Vec{r}$ and the averages are taken over all possible centers on the $\mathbb{S}^2$ sphere.

It can be shown \cite{bovzivc2019spherical} that the spherical cap variance is connected to the spherical structure factor by
\begin{equation}\label{eq:sc_from_pol}
    \sigma_N^2(\theta) = \frac{N_p}{4}\sum_{\ell = 1}^\infty S(\ell) \frac{[P_{\ell+1}(\cos \theta) - P_{\ell-1}(\cos \theta)]^2}{2\ell+1}.
\end{equation}
\eqref{eq:sc_from_pol} implies a vanishing spherical cap variance for $\theta = \pi/2$ for distributions of points symmetric under parity (such as the distributions generated using the LP method).

Moreover, it was suggested by Božič and Čopar that the spherical cap variance behaves as

\begin{equation}\label{eq:behaviour_sc}
\sigma_N^2(\theta) \asymp \frac{A_{N_p}}{4}N_p \sin^2 \theta + \frac{B_{N_p}}{4 \sqrt{3}}\sqrt{N_p} + \xi (\theta),
\end{equation}
where $A_{N_p}$ and $B_{N_p}$ are numerical coefficients that identify the presence of hyperuniformity (namely, hyperuniformity is reached for $A_{N_p}$ = 0) and $\xi(\theta)$ is an additional oscillatory contribution, stronger the more crystalline-like the distribution is.

Hyperuniformity can be defined in terms of the limit $\sigma_N^2(\theta)/s(\theta) \to 0$ for increasing $\theta$, where $s(\theta)$ is the surface of a spherical cap with opening angle $\theta$ \cite{meyra2019hyperuniformity}. Of course, such a limit cannot be truly achieved since we are working on a finite-size sphere and, consequently, the requirement becomes that the ratio decreases sufficiently rapidly for $\theta \to \pi/2$.

It is to be noted that both the definitions of hyperuniformity given above rely on long-range, large-scale properties of the system. Hence, we expect that hyperuniformity will not give much information on the local structure of the set of points. Indeed, as shown in \cite{bovzivc2019spherical, meyra2019hyperuniformity}, a huge variety of different distributions satisfy the criteria of hyperuniformity. This result is in agreement with the results we obtained in this work, which are presented in Sec.~\ref{sec:hyper_results}.

\section{Methods}\label{sec:methods}
\subsection{Algorithms}\label{sec:algorithms}

\begin{figure}[t]
     \centering
     \begin{subfigure}[b]{0.35\textwidth}
         \centering
         \includegraphics[width=\textwidth]{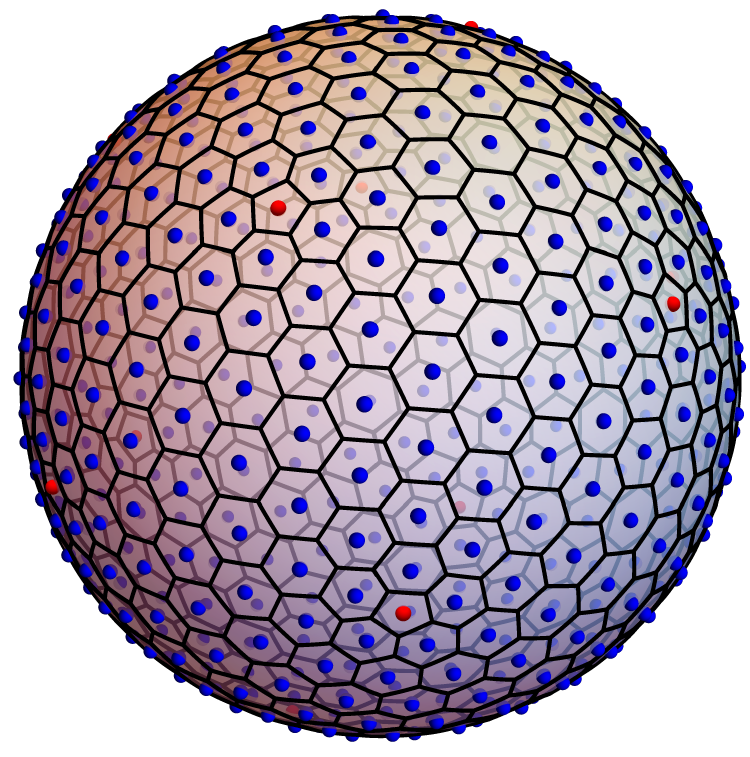}
         \caption{LP, $N_p = 492$.}
         \label{fig:y equals x}
     \end{subfigure}
     \hfill
     \begin{subfigure}[b]{0.35\textwidth}
         \centering
         \includegraphics[width=\textwidth]{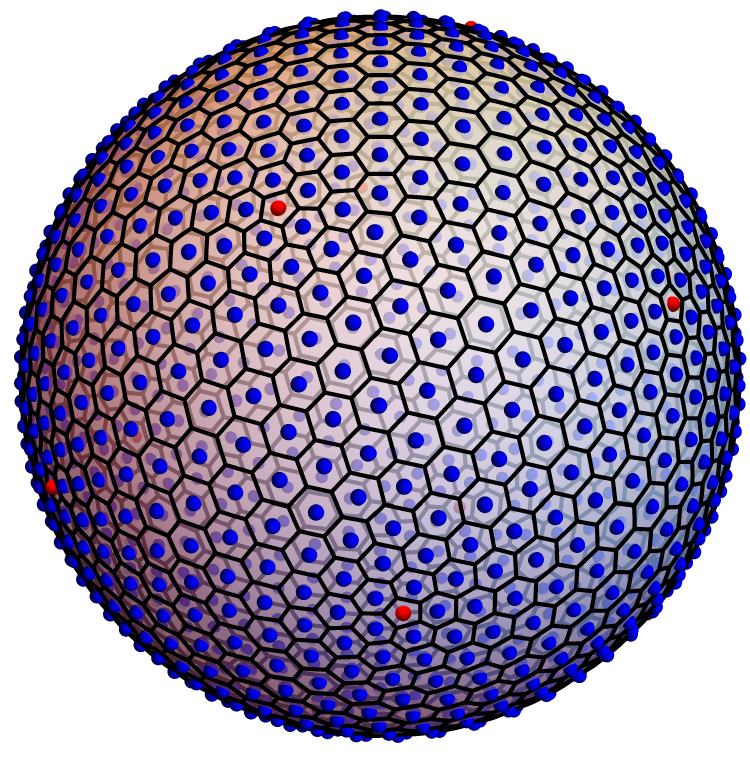}
         \caption{LP, $N_p = 1082$.}
         \label{fig:three sin x}
     \end{subfigure}
     \hfill
     \begin{subfigure}[b]{0.35\textwidth}
         \centering
         \includegraphics[width=\textwidth]{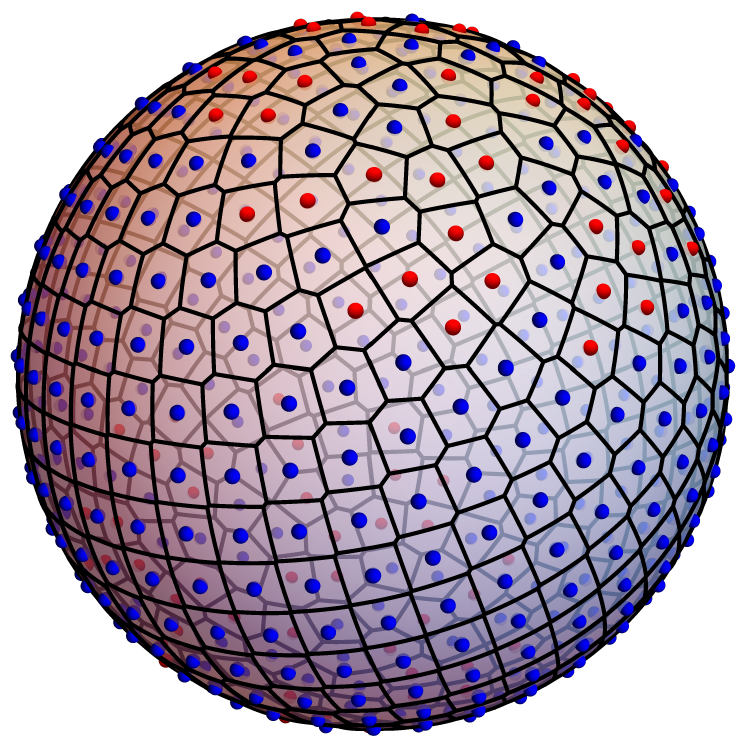}
         \caption{MSP, $N_p = 500$.}
         \label{fig:five over x}
     \end{subfigure}
     \hfill
          \begin{subfigure}[b]{0.35\textwidth}
         \centering
         \includegraphics[width=\textwidth]{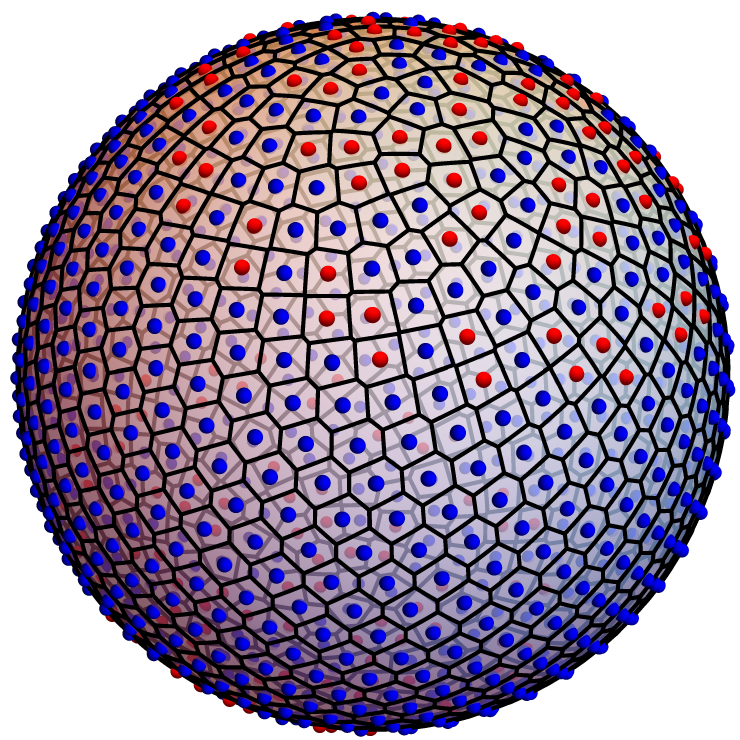}
         \caption{MSP, $N_p = 1000$.}
         \label{fig:five over x}
     \end{subfigure}
        \caption[Examples of point distributions]{Examples of point distributions generated using different algorithms, together with the corresponding Voronoi tessellation. Blue points have six neighbors, while red points are defects with less or more than six neighbors.}
        \label{fig:sfere}   
\end{figure}

\begin{figure}[t]
     \centering
     \begin{subfigure}[b]{0.35\textwidth}
         \centering
         \includegraphics[width=\textwidth]{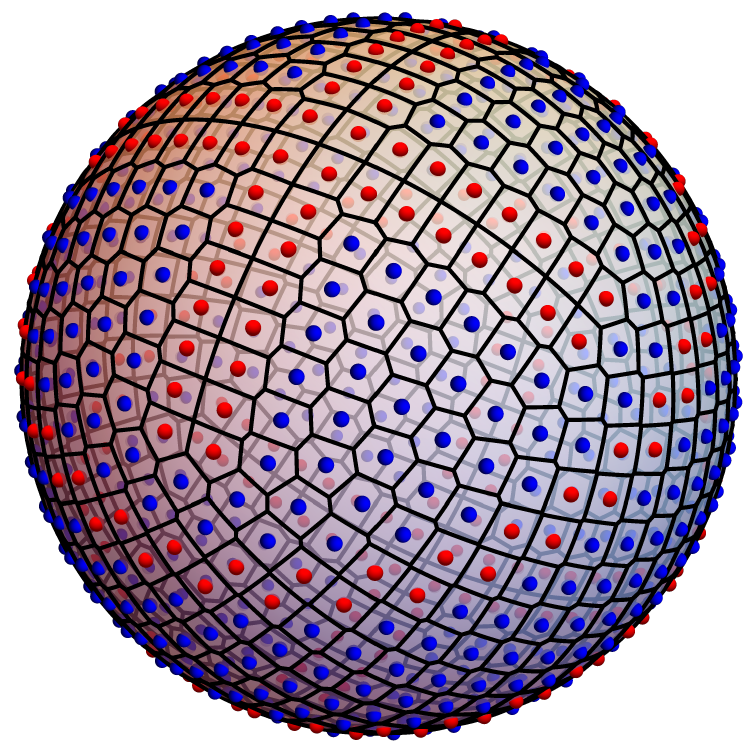}
         \caption{HEALPix, $N_p = 768$.}
         \label{fig:y equals x}
     \end{subfigure}
     \hfill
     \begin{subfigure}[b]{0.35\textwidth}
         \centering
         \includegraphics[width=\textwidth]{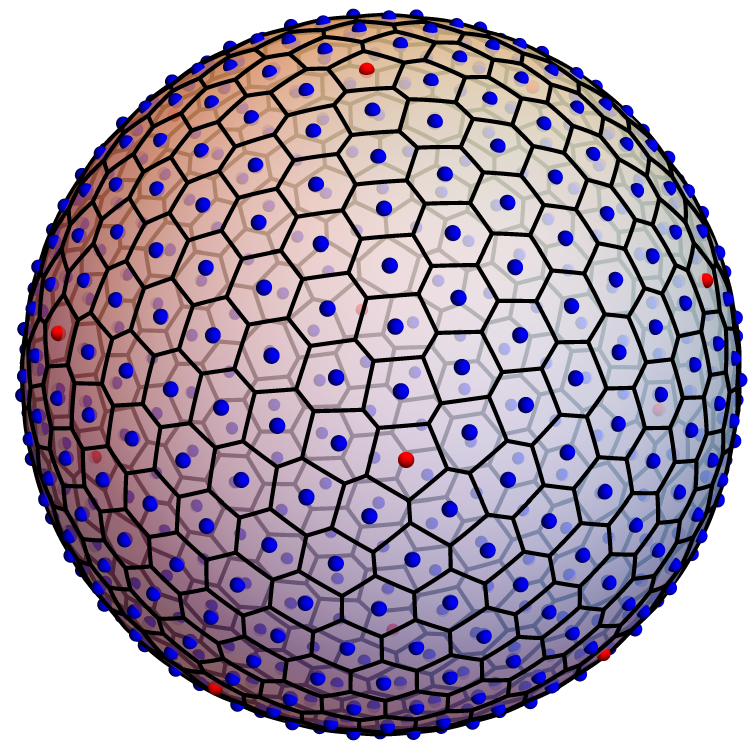}
         \caption{IEAP, $N_p = 492$.}
         \label{fig:three sin x}
     \end{subfigure}
     \hfill
          \begin{subfigure}[b]{0.35\textwidth}
         \centering
         \includegraphics[width=\textwidth]{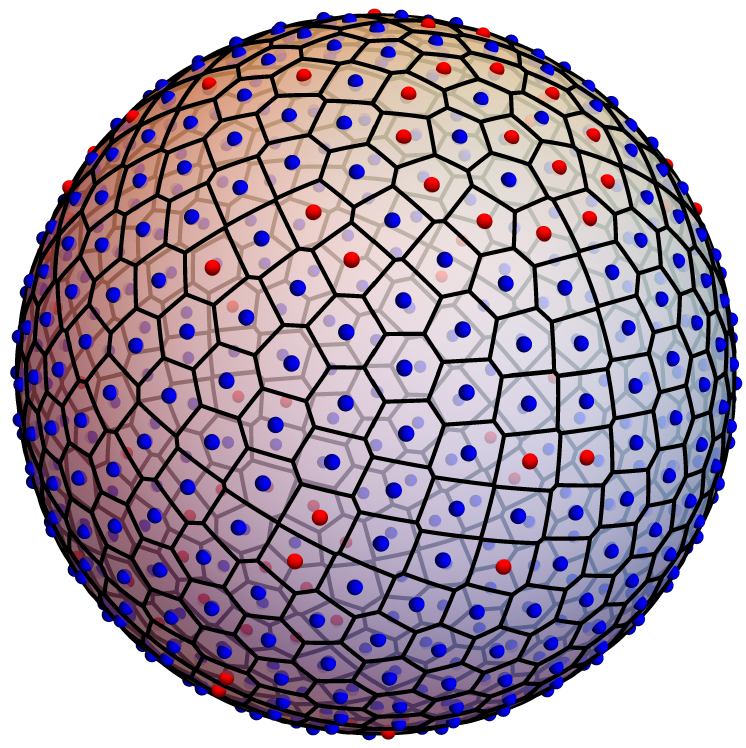}
         \caption{PCS, $N_p = 484$.}
         \label{fig:five over x}
     \end{subfigure}
     \hfill
          \begin{subfigure}[b]{0.35\textwidth}
         \centering
         \includegraphics[width=\textwidth]{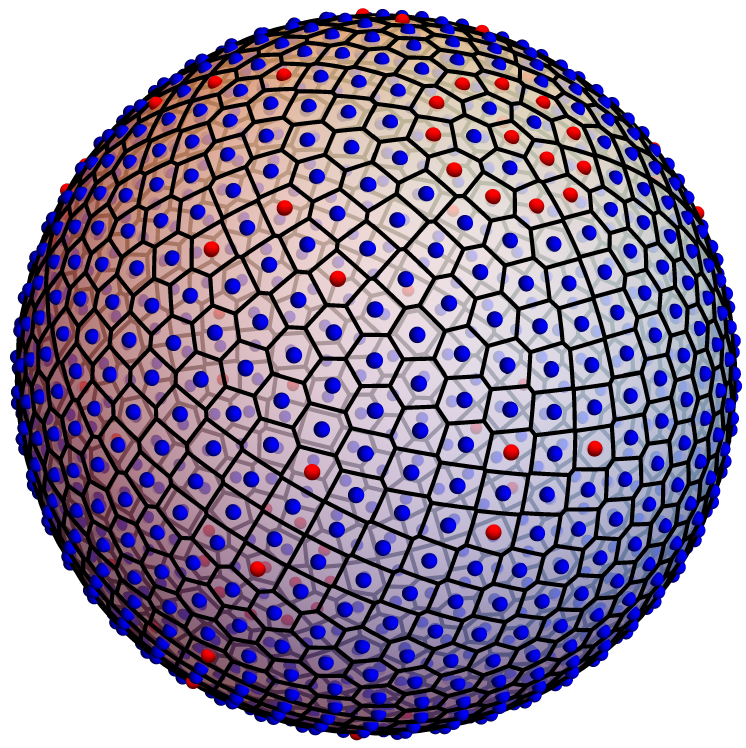}
         \caption{PCS, $N_p = 992$.}
         \label{fig:five over x}
     \end{subfigure}
        \caption[More examples of point distributions]{More examples of point distributions generated using different algorithms, together with the corresponding Voronoi tessellation. As in Fig.~\ref{fig:sfere}, blue points have six neighbors, while red points are defects with fewer or more than six neighbors.}
        \label{fig:sfere2}  
\end{figure}

The procedures we have considered to generate points uniformly on the unitary sphere are listed below.
\begin{enumerate}
    \item \textbf{Lattice Points (LP)} method \cite{altschuler1997possible,katanforoush2003distributing} consisting in the following steps:
    \begin{itemize}
        \item first, an icosahedron inscribed in the sphere is considered;
        \item then, points are added on a face of the icosahedron by mapping the face to an equilateral triangle in the complex plane $\mathbb{C}$ with vertices at 0, $m +n \zeta$ and $m\zeta + n\zeta^2$, where $\zeta \equiv e^{\frac{i\pi}{3}}$ and $m,n$ non-negative integers such that $m \geq n$;
        \item points on the complex plane of coordinates $k + l \zeta$, with $k$ and $l$ integers, which lay inside or on the boundary of the triangle are added to the face;
        \item points on the other faces of the icosahedron are obtained through a similar mapping and appropriate rotations of the points to match them on the edges;
        \item the final distribution of points is obtained by projecting the points from the faces of the icosahedron onto the sphere. 
    \end{itemize}
    The number of points $N_p$ generated for a given pair $m$, $n$ is $N_p = 10(m^2 +n^2 +mn) + 2$;
    
    \item \textbf{Mathematica's \texttt{SpherePoints} (MSP)} function \cite{reference.wolfram_2022_spherepoints}, which takes as input an integer $N_p$ and gives as output an approximately uniform distribution of $N_p$ points on the sphere;
    
    \item \textbf{Polar Coordinates Subdivision (PCS)} method \cite{katanforoush2003distributing}. Denoting by $\varphi$, $\theta$ the polar coordinates on the sphere, the angles $\varphi_j = \pi(\frac{j}{n} - \frac{1}{2})$ ($n$ is a parameter that fixes the number of points generated) are considered. For each $\varphi_j$, $n_j = \lfloor\frac{1}{2} +\sqrt{3}n \cos \varphi_j \rfloor$ equally spaced points are placed at angles $\theta_k = \frac{2\pi}{n_j}k$. One can add the two poles (we did not). Finally, a shift on alternate latitudes is introduced to symmetrize the distribution. This method can be slightly modified to increase the number of $N_p$ available, but we did not consider this modified version in this work;
    
    \item the \textbf{HEALPix} package,\footnote{https://healpix.sourceforge.io/.} a structure to obtain iso-latitude, equal-area pixels on the sphere organized in a hierarchical way, which is widely used in the analysis of astronomical data \cite{Gorski_2005}. HEALPix creates pixels by subsequent subdivision of twelve original pixels. At each step, each pixel is divided into four sub-pixels. At the end of the procedure, a discretization of the sphere can be obtained by taking each pixel's center;
    
    \item \textbf{Icosahedron-based Equal Area Pixelization (IEAP)} method \cite{Tegmark_1996}, which consists of the following steps:
    \begin{itemize}
        \item first, the faces of an icosahedron with a regular triangular grid are pixeled. Each pixel is identified by a point which represents its center;
        \item the points obtained in the previous step are projected onto the unit sphere;
        \item the points are shifted around slightly so that each pixel has an equal area.
    \end{itemize}
\end{enumerate}

Eventually, a Gradient Descent (GD) procedure can be applied to the sets of points obtained by the above algorithms to further improve the final result. We will discuss some possible choices of potentials to perform the GD optimization and their effects.

Examples of point distribution obtained using the five algorithms described above are shown in Fig. \ref{fig:sfere} and \ref{fig:sfere2}.

\subsection{Potentials for the GD}\label{sec:potentials}

To study the effects that different potentials have on the GD procedure, we consider 4 kinds of potentials:
\begin{enumerate}
    \item Power-law potentials:
    \begin{equation}
        V_1(r; \alpha) \equiv \frac{1}{r^\alpha};
    \end{equation}
    \item Power-law potentials with an exponential cutoff:
    \begin{equation}
        V_2(r; \alpha) \equiv \frac{e^{-\frac{r}{r_0}}}{r^\alpha};
    \end{equation}
    \item Scaled power-law potentials:
    \begin{equation}
        V_3(r; \alpha) \equiv \Big (\frac{r_0}{r} \Big )^\alpha;
    \end{equation}
    \item Scaled power-law potentials with an exponential cutoff:
    \begin{equation}
        V_4(r; \alpha) \equiv \Big (\frac{r_0}{r} \Big )^\alpha e^{-\frac{r}{r_0}}.
    \end{equation}
\end{enumerate}
The characteristic length scale $r_0$ has been chosen as the average distance between two nearest neighbors, estimated assuming a hexagonal lattice local structure for the distribution:
\begin{equation}\label{eq:char_len}
    r_0 = \sqrt{\frac{16\eta}{N_p}},
\end{equation}
$\eta$ being the packing density of the 2D hexagonal lattice, $\eta = \frac{\pi}{\sqrt{12}} \approx 0.91$. The correctness of expression in \eqref{eq:char_len} has been verified by looking at both non-optimized and optimized configurations (see Fig.~\ref{fig:mean}).

\section{Results}\label{sec:results}

Now we present the results obtained using different algorithms and optimizing potentials. Distances are always Euclidean, given that the difference with the geodesic distances is very small when considering nearby points and can be safely neglected.  

\subsection{Comparison with the random case}\label{sec:comparison_random}

\begin{figure}[t]
     \centering
     \begin{subfigure}[h]{0.5\textwidth}
         \centering
         \includegraphics[width=\textwidth]{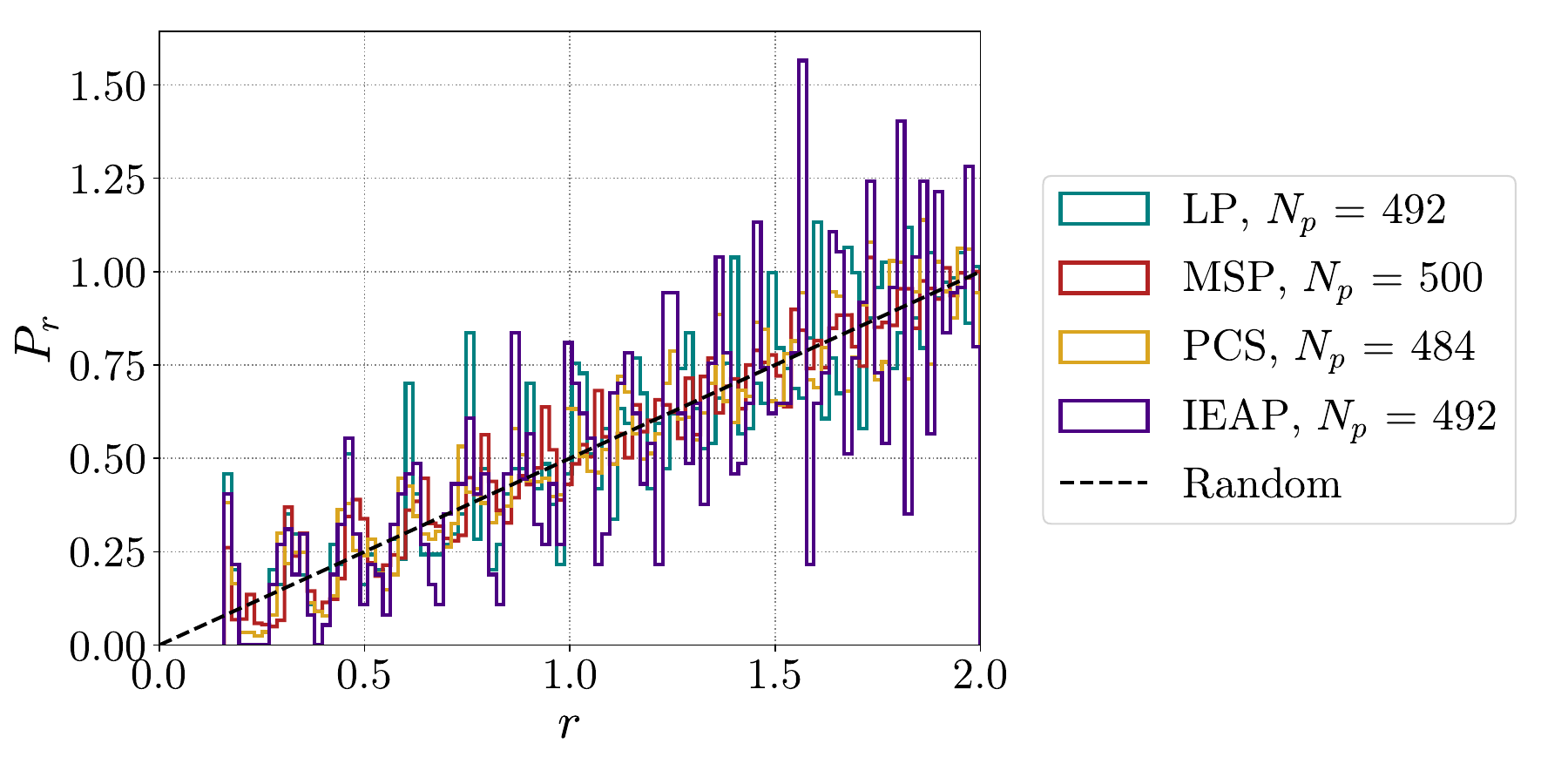}
         \caption{Distances.}
         \label{fig:g_confronto}
     \end{subfigure}
     \begin{subfigure}[h]{0.5\textwidth}
         \centering
         \includegraphics[width=\textwidth]{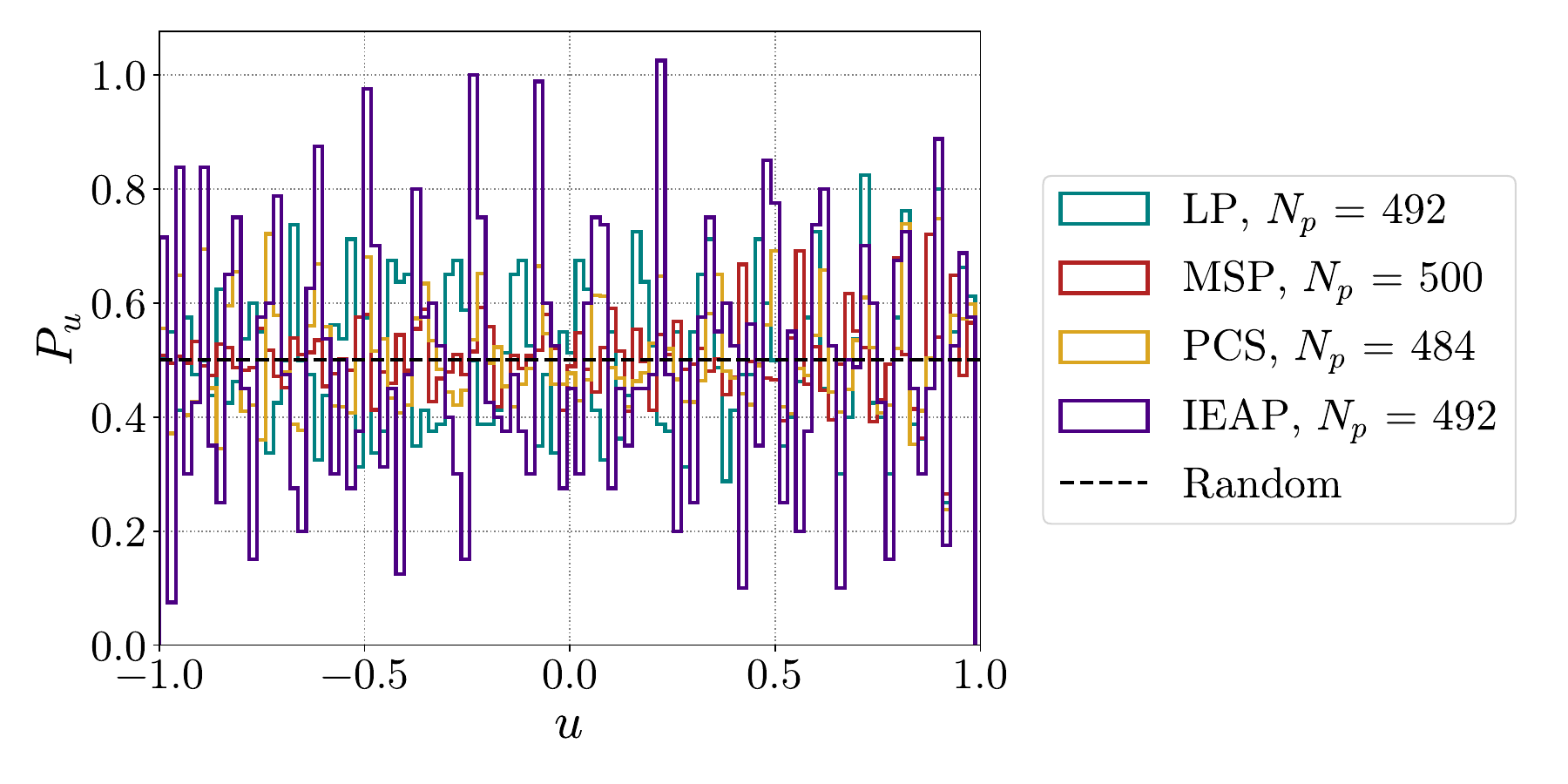}
         \caption{Scalar products.}
         \label{fig:scalar_confronto}
     \end{subfigure}
    \begin{subfigure}[h]{0.5\textwidth}
         \centering
         \includegraphics[width=\textwidth]{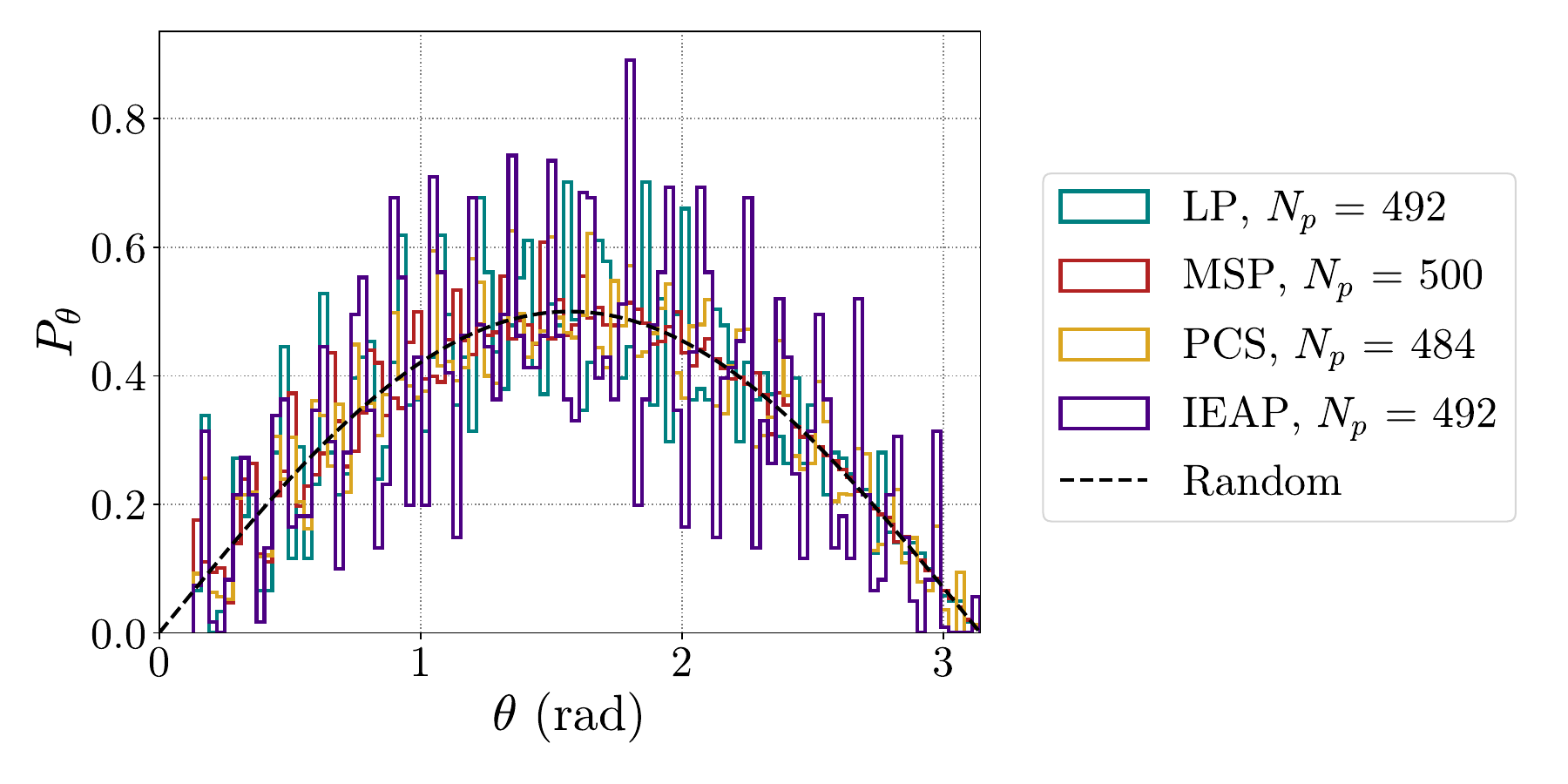}
         \caption{Angles.}
         \label{fig:geodesic_confronto}
     \end{subfigure}
        \caption[Probability distributions of distances, scalar products and angles (LP, MSP, PCS and IEAP methods)]{Comparison between the probability distributions of distances, scalar products and angles between points obtained via LP, MSP, PCS and IEAP methods and the theoretical prediction in the case of points placed uniformly at random on the sphere.}
\end{figure}

First, we compare the probability distribution of distances found using the algorithms LP, MSP, PCS and IEAP with the theoretical prediction for points distributed at random according to a uniform distribution. For a unit hypersphere in dimension $m$, the probability distribution of distances, in the random case, is given by
\begin{equation}
  P_r(r) = \frac{r^{m-2}(1- \frac{r^2}{4})^{\frac{m-3}{2}}}{A_m}, \rlap{ \; \; \; \; \;          $0 \leq r \leq 2$}
\end{equation}
where
\begin{equation}
    A_m \equiv \int_0^\pi d \theta (\sin \theta)^{m-2},
\end{equation}
which simply reduces to $P_\text{rand}(r) = r/2$ in our $m = 3$ case. The comparison is shown in Fig.~\ref{fig:g_confronto}. For all algorithms, fluctuations from the linear behavior of the random case are present, signaling the presence of structure. In particular, these fluctuations appear to be bigger for the LP algorithm at smaller $r$ (the region we are most interested in when looking for crystalline-like order) and for the IEAP algorithm at larger $r$.
Similarly, in Fig.~\ref{fig:scalar_confronto} and \ref{fig:geodesic_confronto}, the scalar products and the angles between points have been considered. For random points on the sphere we have
\begin{equation}
    P_u(u) = \frac{(1-u^2)^\frac{m-3}{2}}{A_m}, \rlap{\; \; \; \; \; \;       $-1 \leq u \leq 1$}
\end{equation}
for scalar products and
\begin{equation}
        P_\theta(\theta) = \frac{(\sin \theta)^{m-2}}{A_m}, \rlap{ \; \; \; \; \; \;  \, \,     $0 \leq u \leq \pi$}
\end{equation}
for angles, which reduce to $P_u(u) = 1/2$ and $P_\theta(\theta) = \sin (\theta)/2$ in the $m = 3$ case we are considering.

\begin{figure}[t]
    \centering
    \includegraphics[width=0.6\textwidth]{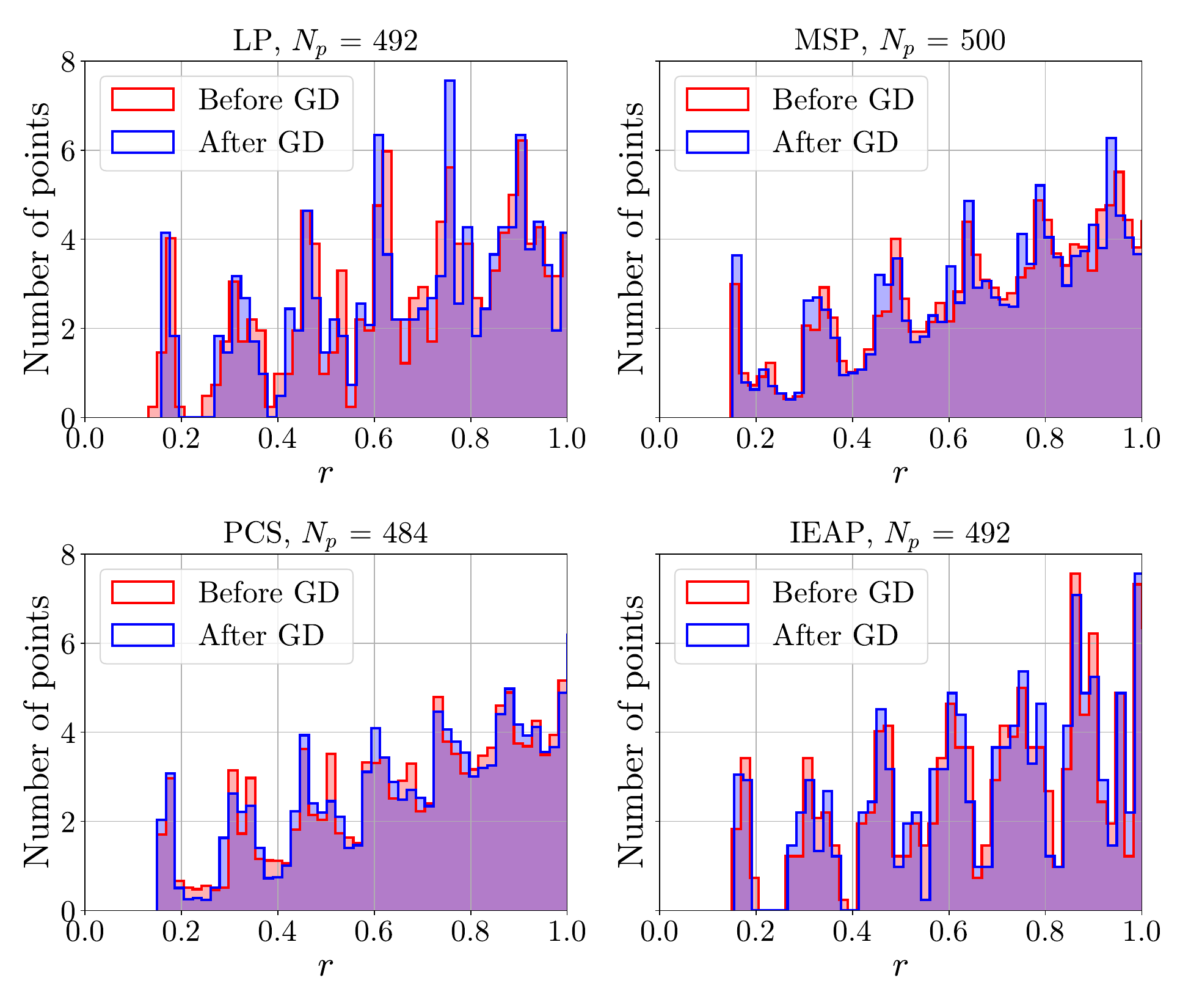}
    \caption{Histograms of the number of points at distance $r$. LP, MSP, PCS and IEAP methods are considered. Red and blue bars represent the distribution before and after Coulomb GD optimization, respectively. The first peak is well separated for LP and IEAP, but not for MPS and PCS. Notably, GD makes the first peak sharper.}
    \label{fig:g}
\end{figure}

\begin{figure}[!h]
    \centering
    \includegraphics[width=0.6\textwidth]{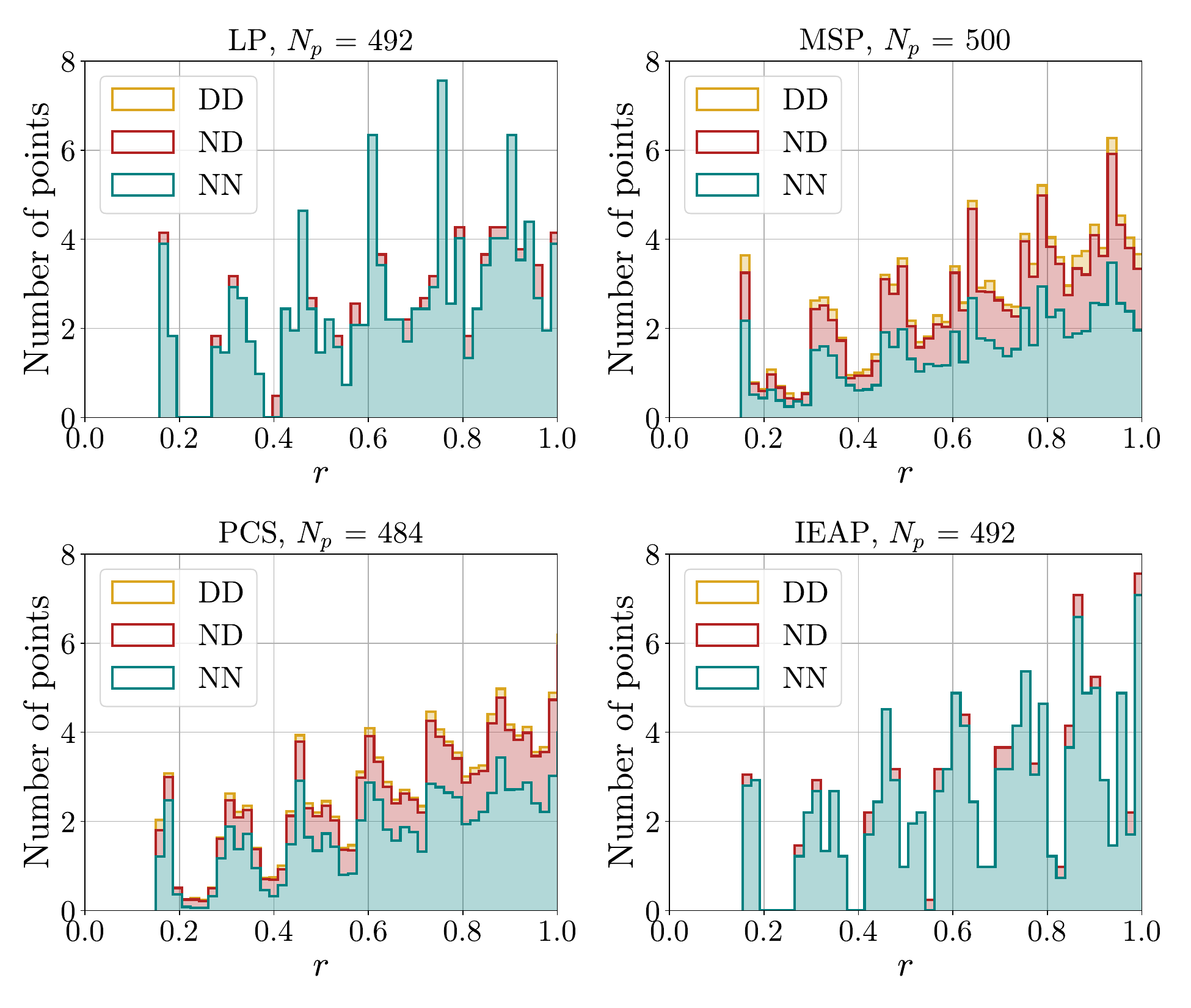}
    \caption{Histograms of the number of points at distance $r$ for LP, MSP, PCS and IEAP methods, after Coulomb GD optimization. Bars are placed on top of each other, so the total distribution of points is given by the highest contour. Different colors identify different kinds of points interacting. \textit{Bottom, teal}: normal-normal (NN). \textit{Middle, red}: normal-defect (ND). \textit{Top, yellow}: defect-defect (DD). In the LP and IEAP methods, the number of defects is so small compared to the number of normal points that defect-defect interactions do not appear visually in the figure.}
    \label{fig:g_wtypes}
\end{figure}

\subsection{Distributions of distances}\label{sec:dist_distr}

In Fig.~\ref{fig:g},  the histograms of the average number of points at a distance $r$ from a given point, obtained through algorithms LP, MSP, PCS and IEAP are shown. In the LP and IEAP cases, the first peak (corresponding to nearest neighbors) can be easily identified, whereas in the MSP and PCS the first peak mixes with the second one. The HEALPix distribution has a behavior qualitatively similar to that of PCS and MSP, with a first peak not well separated from the second one. Moreover, it is a rather sparse algorithm in this range of $N_p$. That is, there is only one configuration with a number of points between 200 and 3000. This makes HEALPix unsuited for analyzing the effects of discretization, e.g.\ to study finite size effects. For these two reasons, it was not considered further.

Fig.~\ref{fig:g_wtypes} shows the histograms obtained using the different methods (followed by a Coulomb GD) while the type of points (normal points with six neighbors or defects) is taken into account. PCS and MSP produce a huge number of defects.

It is clear that the first peak of the distribution is the one that carries the most information about the local structure of the distribution, so we will analyze it more carefully. Before doing so, however, we present the results obtained testing hyperuniformity criteria on the different distributions.


\subsection{Hyperuniformity results}\label{sec:hyper_results}

\begin{figure}[t]
\centering  
    \begin{subfigure}[b]{0.4\textwidth}
    \includegraphics[width=\textwidth]{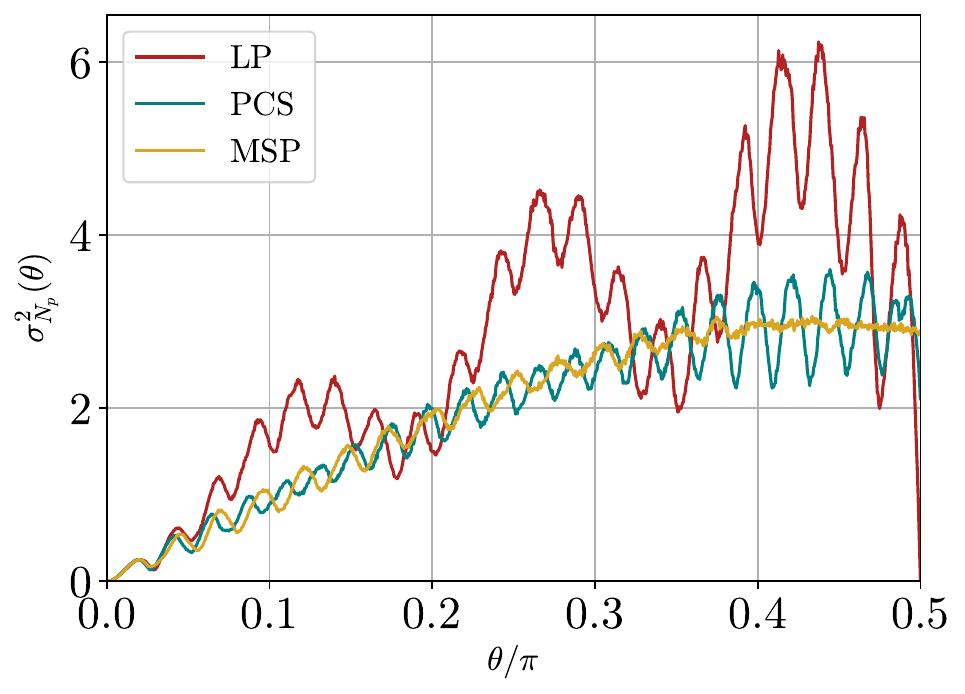}
    \caption{Spherical cap.}
    \label{fig:cap}
    \end{subfigure}
    \hspace{1cm}
    \begin{subfigure}[b]{0.4\textwidth}
    \includegraphics[width=\textwidth]{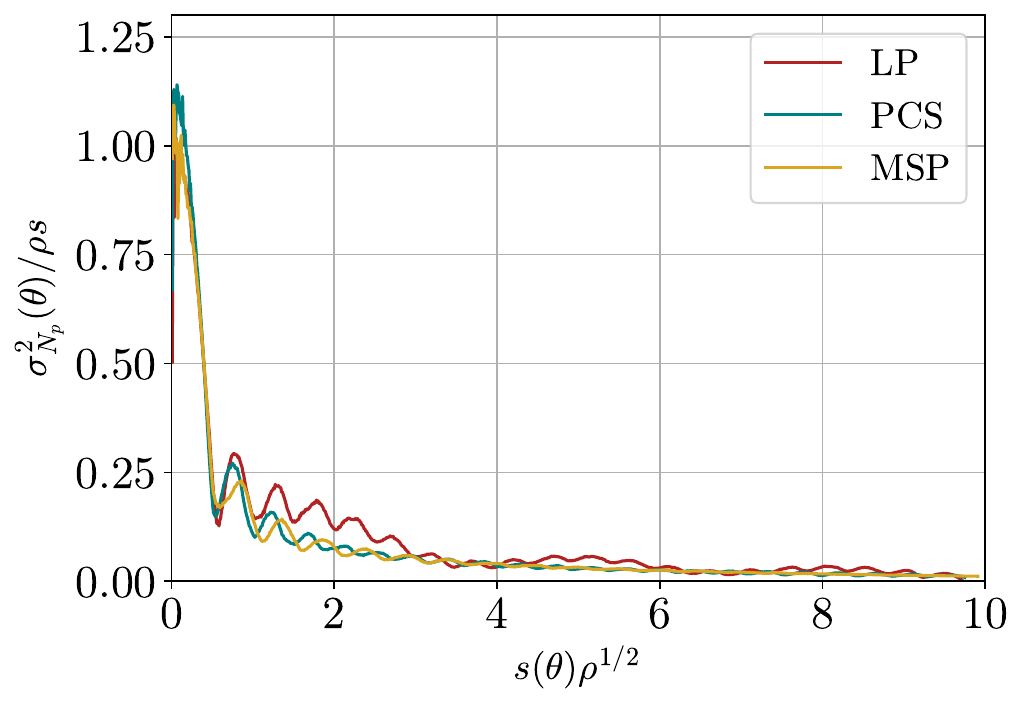}
    \caption{Normalized spherical cap.}
    \label{fig:cap_norm}
    \end{subfigure}
    \caption[Spherical cap variance for LP, PCS and MSP]{(a) spherical cap variance and (b) spherical cap variance normalized by spherical cap area, as functions of the angle, for different point distribution: LP ($N_p = 482$, $V_3$ GD optimization with $\alpha = 1$), MSP ($N_p = 500$) and PCS ($N_p = 484$). $\rho$ is the density of points, i.e. $N/4\pi$. The spherical cap variance was estimated taking the average over 20000 random centers of the cap for every angle. Notice the vanishing $\sigma_N^2$ for $\theta = \pi/2$ for the LP method. This is a consequence of the symmetry of the distribution under space inversion, which implies a zero structure factor for odd $\ell$, which in turn means that only odd Legendre polynomials enter in the sum of \eqref{eq:sc_from_pol}.}
    \label{fig:scap_together}
\end{figure}

\begin{figure}[t]
    \centering
\includegraphics[width=0.5\textwidth]{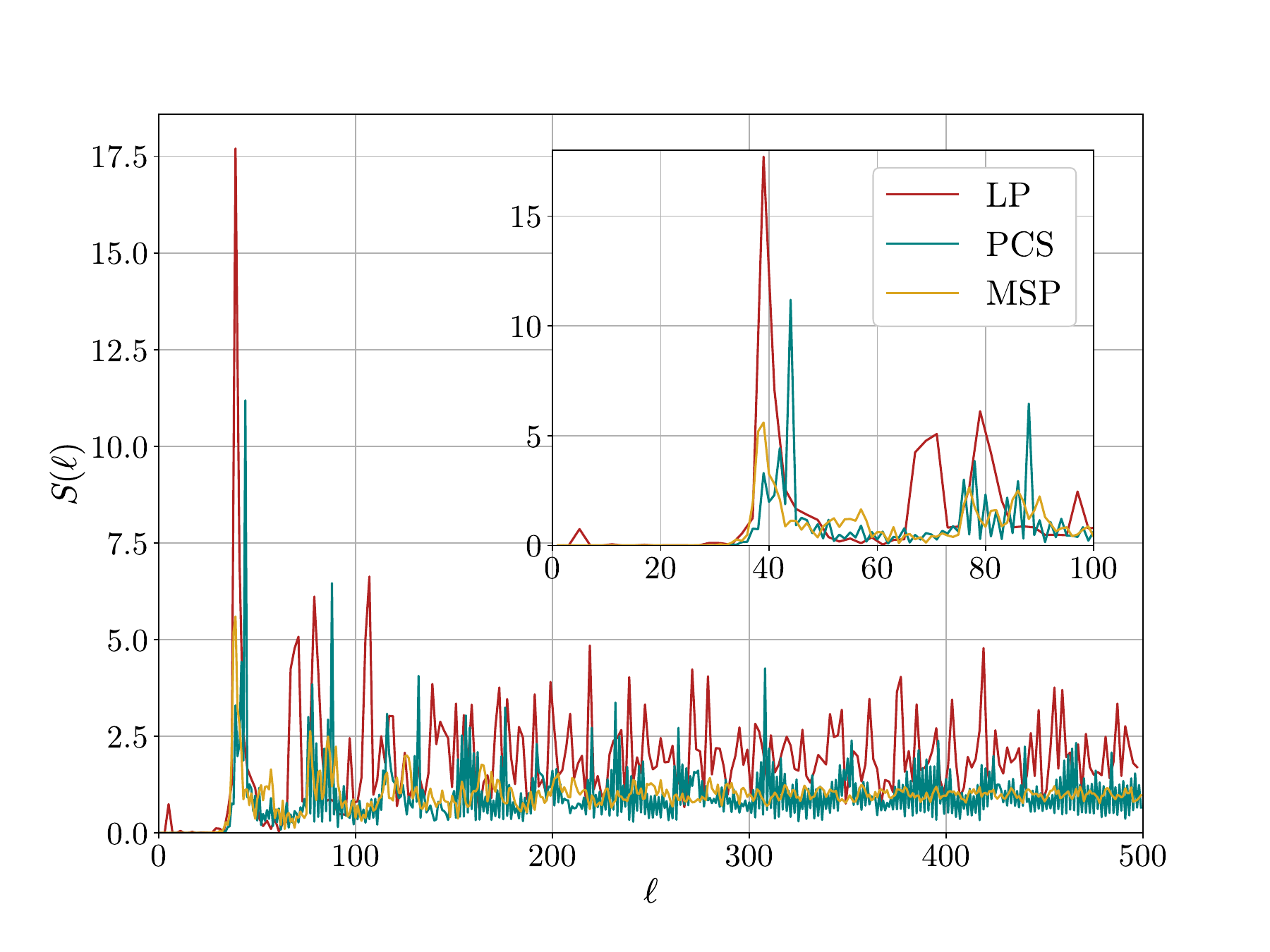}
    \caption[Structure factor for LP, PCS and MSP]{\textit{Main plot}: structure factor as a function of the wave number $\ell$ for different point distribution: LP ($N_p = 482$, $V_3$ GD optimization with $\alpha = 1$), MSP ($N_p = 500$) and PCS ($N_p = 484$). \textit{Inset}: zoom in on the $\ell \leq 100$ region. For the LP method, only even $\ell$ are considered, since the structure factor vanishes at odd $\ell$ for symmetry reasons. }
    \label{fig:structure}
\end{figure}

We calculated both the structure factor and the spherical cap variance for three distributions of points obtained through the LP ($N_p = 482$), MSP ($N_p = 500$) and PCS ($N_p = 484$) methods. The spherical cap variance is presented in Fig.~\ref{fig:cap} and the spherical cap variance normalized using the surface of the cap is shown in Fig. \ref{fig:cap_norm}. It is clear that, in all three cases, the ratio $\sigma_N^2(\theta)/s(\theta)$ goes to zero for increasing $\theta$, thus suggesting that all the distributions of points satisfy hyperuniformity. This is further confirmed by the behavior of the structure factor, shown in Fig.~\ref{fig:structure}. Indeed, a vanishing $S(\ell)$ is observed at small $\ell$, together with a well-defined first peak.

We notice that in Fig.~\ref{fig:cap} the behaviors of the three distributions appear to follow \eqref{eq:behaviour_sc}. The LP configuration, however, shows bigger fluctuations and actually seems to possess two natural frequencies, compared to only one as in the MSP and PCS cases. Since the oscillatory term $\xi(\theta)$ in \eqref{eq:behaviour_sc} is connected to crystalline-like ordering (it was observed that random distributions of points do not fluctuate at all), this might be additional proof that LP method indeed produces more ordered structures than the other two methods. 

As previously stated, hyperuniformity does not seem to be a strong enough criterion to identify spatially uniform distributions. It is, however, interesting to notice that, while PCS and MSP give distributions that reach a unitary structure factor relatively quickly, it takes LP much longer to reach the asymptotic regime.\footnote{The structure factor must go to 1 for $\ell \to \infty$. Indeed, $S = 1$ for a completely random distribution of points.} Since large $\ell$ corresponds to large wave vectors in the Euclidean case, it seems reasonable to consider this slower decay as an indication of the presence of local structure.

\subsection{Study of the first peak}\label{sec:studyfirstpeak}

In order to study the uniformity of the distribution, we studied the first peak of the $g(r)$, which, apart from a normalization factor, coincides with the histograms previously obtained.

First of all, we can introduce a measure of how well-separated the first peak is, in order to support the visual evidence presented in Fig. \ref{fig:g} and \ref{fig:g_wtypes}.
 We notice that the rightmost point in the first peak should, in the icosadeltahedral case,  correspond to the $n^* = 6(N-2)$-th point of the distribution. Moreover, it should be well separated from the points on the right and very close to the points on the left. So we can consider the ratio
\begin{equation}
    R \equiv \sum_{i = n^*-\nu}^{n^*-1} (r_{n^*} - r_i)/ \sum_{i = n^*+1}^{n^*+\nu} (r_i - r_{n^*})
\end{equation}
where $r_i$ is the $i$-th smallest distance between 2 points and $\nu$ is a parameter chosen to smooth the behaviour. 
As an alternative, we can also consider the ratio $(r_{n^* + \nu} - r_{n^*}) / (r_{n^*} - r_{n^*-\nu})$. In the well-separated case, $R$ should be close to 0. In Tab. \ref{tab:R_alt} and \ref{tab:R_math} the values of $R$ for different choices of parameters in the LP and MSP cases are presented. As expected, in the case of a well-defined peak $R\approx 0$, while in the case of a crossover behaviour $R \approx 1$.

\begin{table}[t]
\begin{subtable}[b]{0.4\textwidth}
\centering
\begin{tabular}{|c|c|c|c|}
\hline
$m$ & $n$ & $N_p$ & $R$      \\ \hline \hline
3   & 2   & 192 & 1.93$\cdot 10^{-15}$ \\ \hline
3   & 3   & 272 & 2.81$\cdot 10^{-15}$ \\ \hline
4   & 2   & 282 & 2.47$\cdot 10^{-15}$ \\ \hline
4   & 3   & 372 & 2.41$\cdot 10^{-15}$ \\ \hline
\end{tabular}
\caption{LP}
\label{tab:R_alt}
\end{subtable}
\begin{subtable}[b]{0.4\textwidth}
\centering
\begin{tabular}{|c|c|}
\hline
$N_p$ & $R$ \\ \hline \hline
48  & 0.868 \\ \hline
96  & 1.169 \\ \hline
192 & 0.896 \\ \hline
357 & 0.867 \\ \hline
\end{tabular}
\caption{MSP}
\label{tab:R_math}
\end{subtable}
\caption{Values of $R$ obtained using LP and MSP algorithm for different choices of parameters. For the smoothing parameter we take $\nu = 100$. }
\label{tab:R}
\end{table}

From the discussion made in Sec.~\ref{sec:defects} and Sec.~\ref{sec:meas_unif}, it is clear that, of all the algorithms considered, only those that produce icosadeltahedral configurations, or that at least keep the first peaks well-separated, have a good distribution of points. Of all the methods considered in the previous section, only LP and IEAP satisfy this criterion. The latter produces configurations for fewer values of the number of points\footnote{Indeed, LP generates configurations with $N_p = 10(m^2 +n^2 + mn) +2$, for $m, n \in \mathbb{N}$, as previously mentioned, while IEAP produces configurations with $N_p = 40n(n-1) +12$, $n \in \mathbb{N}$.} and has a slightly larger value, after the GD, of the STD/mean observable (as described later). We therefore focus on the LP method. Since we found a good option for the starting algorithm, the next logical step is to consider different potentials for the GD procedure and see how things change.

\begin{figure}[t]
    \centering
    \includegraphics[width=0.75\textwidth]{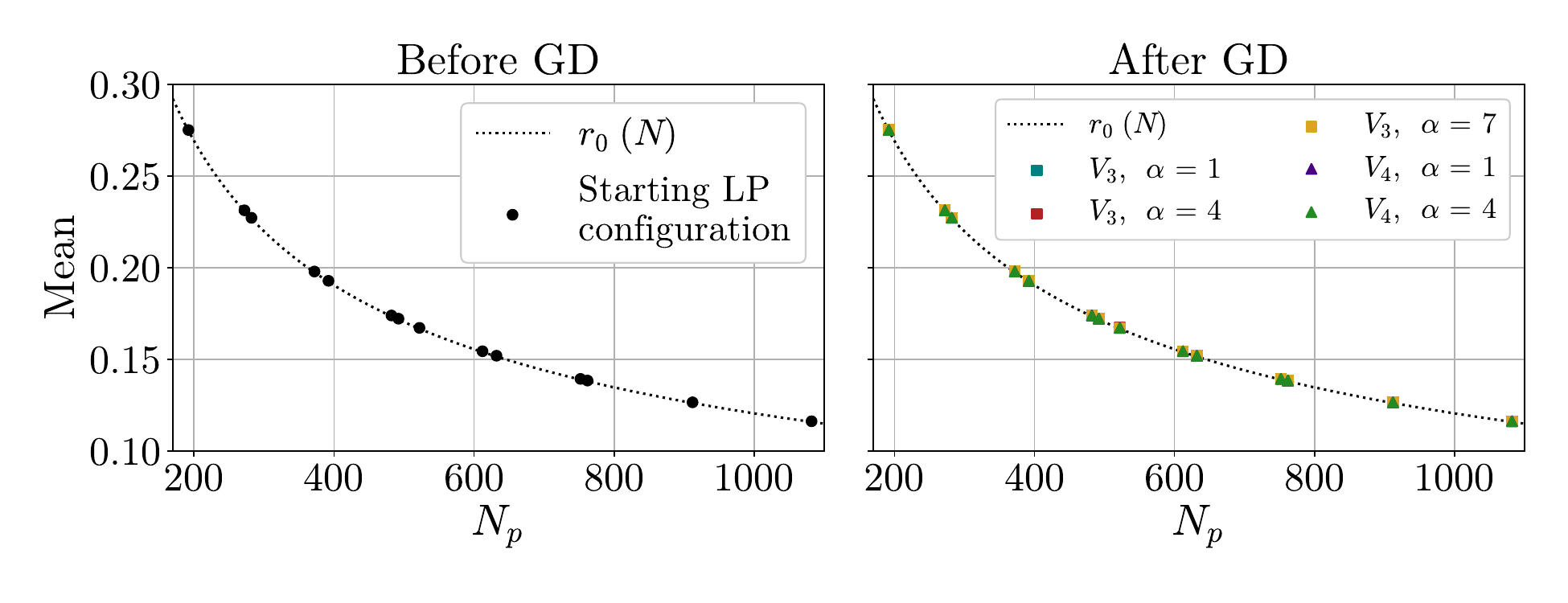}
    \caption[Mean distance inside the first peak]{Mean distance inside the first peak as a function of the number $N_p$ of points placed on the sphere, both for the original configuration (left) and for the configuration optimized by GD (right). In the latter case, the $V_3$ and $V_4$ potentials are considered. Dotted line is given by \eqref{eq:char_len}.}
    \label{fig:mean}
\end{figure}
\begin{figure}[t]
     \centering
    \includegraphics[width=0.75\textwidth]{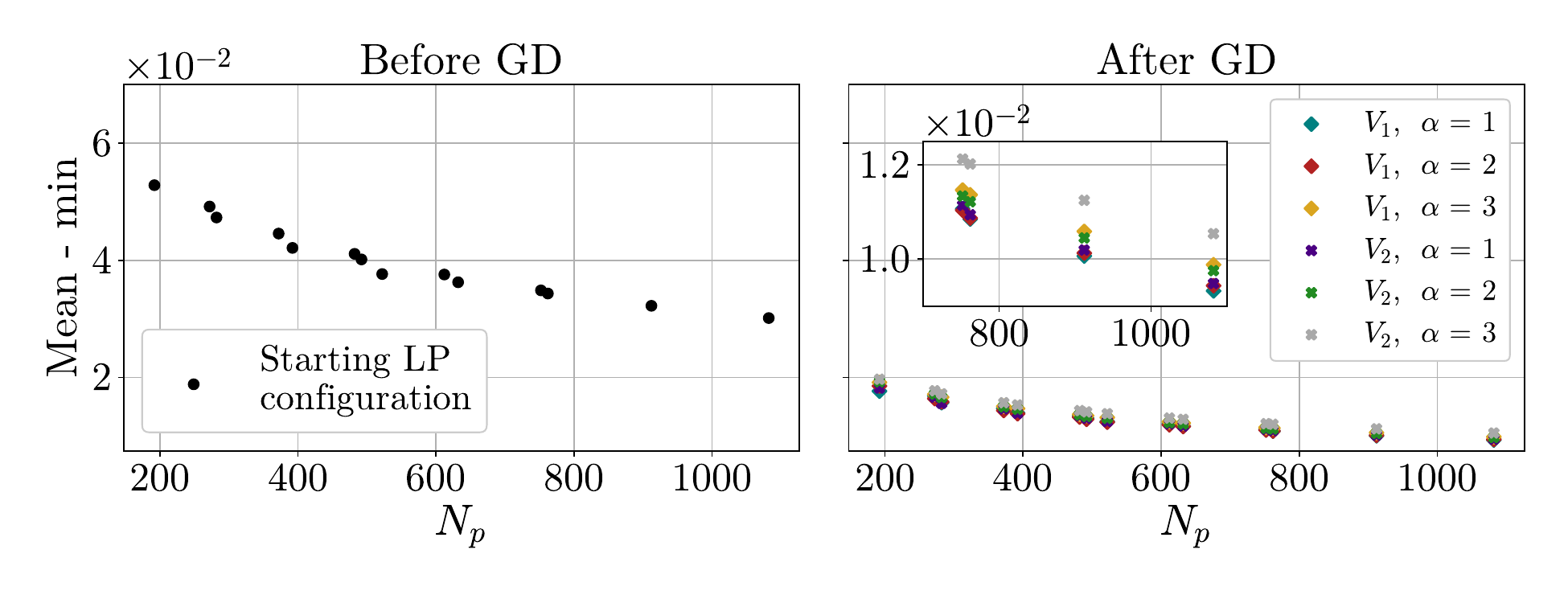}
    \includegraphics[width=0.75\textwidth]{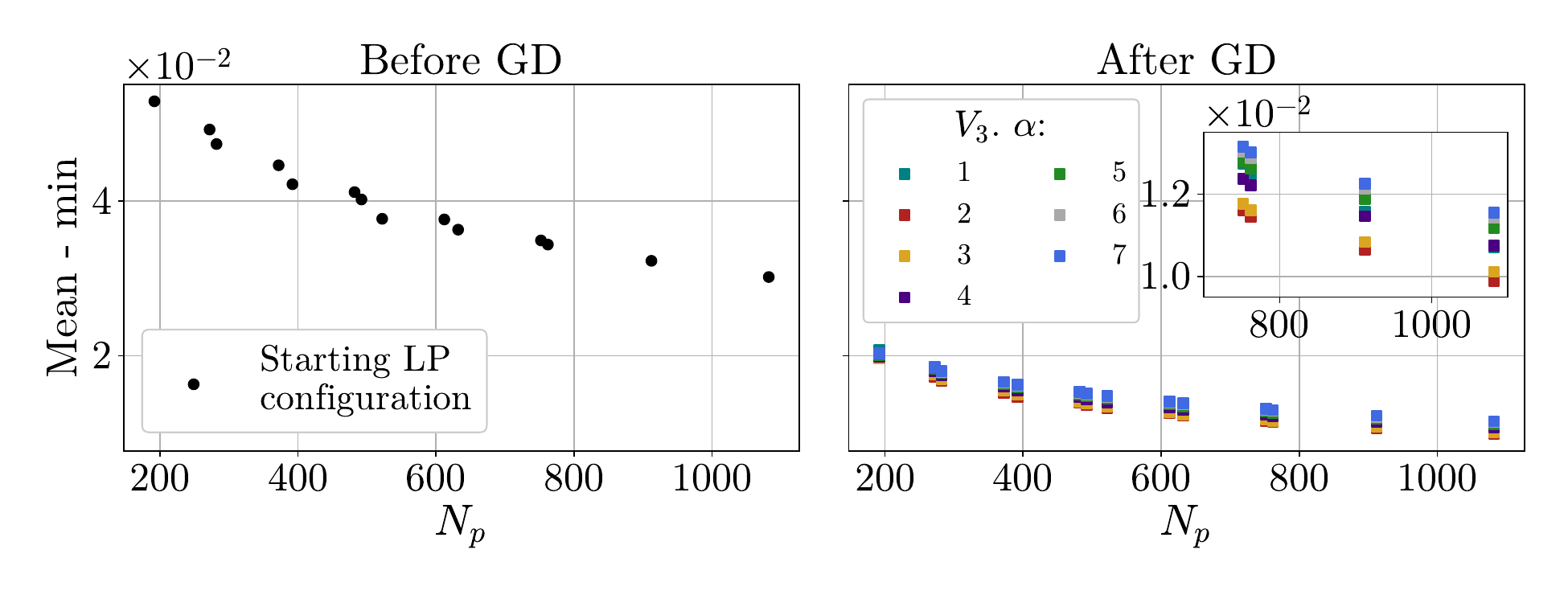}
    \caption[Difference between the mean of the distances inside the first peak and the minimum distance]{Difference between the mean of the distances inside the first peak and the minimum distance as a function of the number $N_p$ of points placed on the sphere, both for the original configuration (left) and for the configuration optimized by GD (right). Insets show zoom-ins of the region at larger values of $N_p$. Top: $V_1$ and $V_2$. Bottom: $V_3$.}
    \label{fig:meanmin}    
\end{figure}

\begin{figure}[t]
    \centering
    \includegraphics[width=0.75\textwidth]{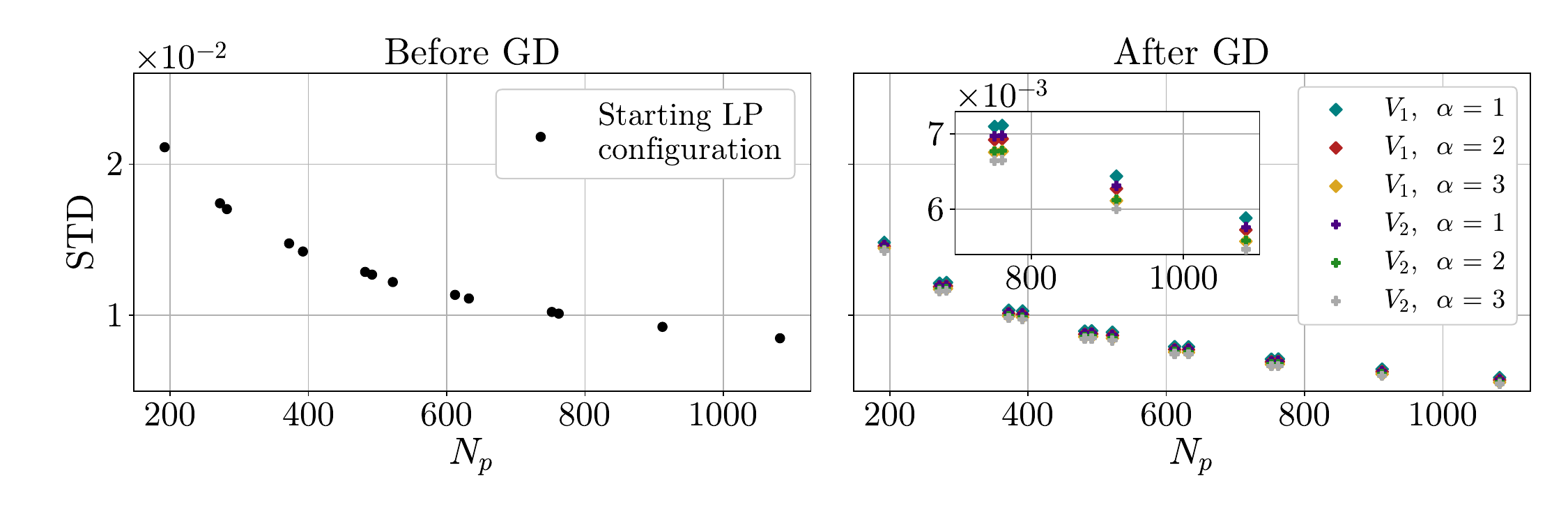}
    \includegraphics[width=0.75\textwidth]{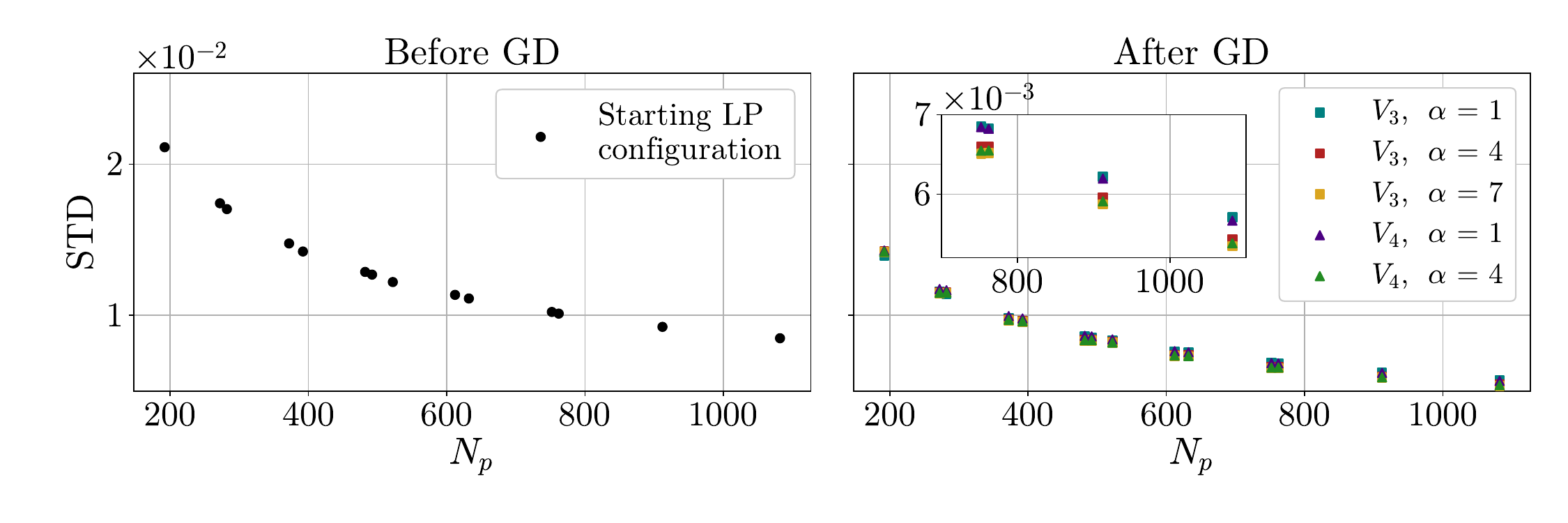}
    \caption{Standard deviation of distances inside the first peak as a function of the number $N_p$ of points placed on the sphere, both for the original configuration (left) and for the configuration optimized by GD (right). Insets show zoom-ins of the region at larger values of $N_p$. Top: $V_1$ and $V_2$. Bottom: $V_3$ and $V_4$.}
    \label{fig:std}    
\end{figure}

\begin{figure}[t]
    \centering
    \includegraphics[width=0.75\textwidth]{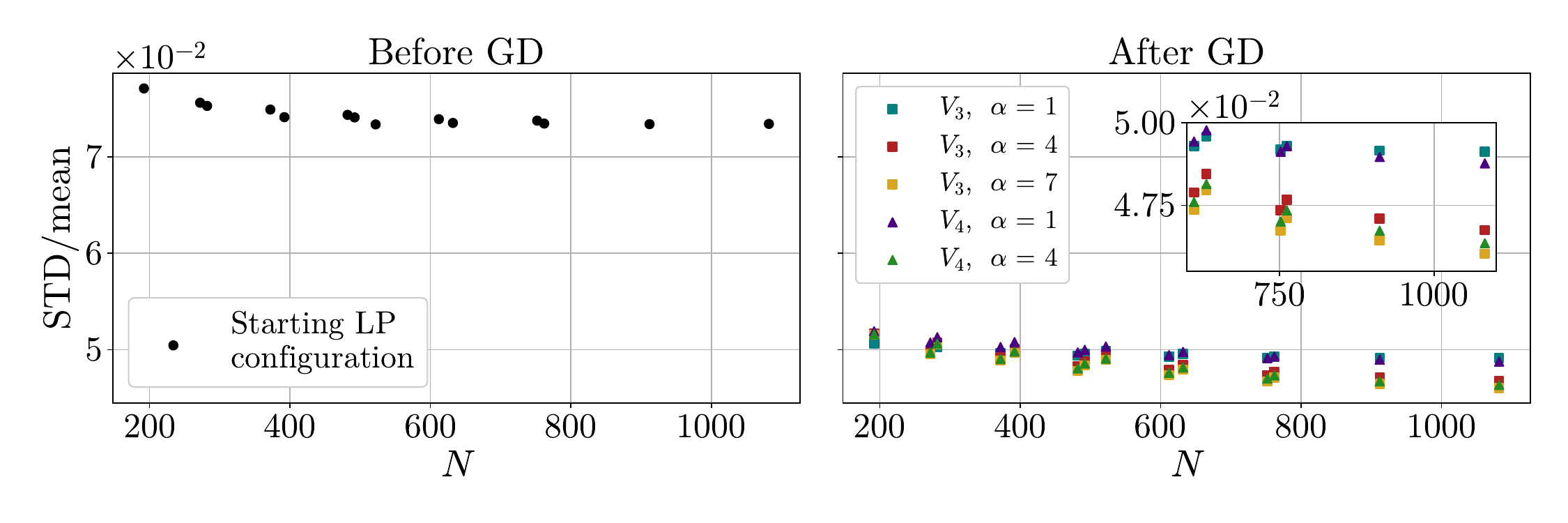}
    \caption{Ratio between the standard deviation and the mean of distances inside the first peak as a function of the number $N_p$ of points placed on the sphere. and $V_4$ potentials are considered. Both graphs share the same $y$ axis.}
    \label{fig:ratio}
\end{figure}

Now we study the effect of optimizing the points positions by minimizing different interaction potentials. We analyzed the first peak of the distribution obtained after a GD optimization, considering its mean, the difference between the mean and the minimum, the standard deviation STD and the ratio STD/mean.
Results are shown in Fig. \ref{fig:mean} to \ref{fig:ratio}.

The mean appears to be largely unaffected by the GD (Fig. \ref{fig:mean}), as can be expected since the total surface of the sphere is fixed. Moreover, it follows the expected curve given by \eqref{eq:char_len}.

The difference between the mean and the minimal value in the first peak is generally a non-monotonic function of $\alpha$, especially in the $V_3$ case (Fig.~\ref{fig:meanmin}). It is however important to remember that this observable, taking into account the minimum distance between two points, measures one extreme of the distribution, thus it is strongly affected by outliers. It is nonetheless interesting to notice that the GD procedure greatly reduces the difference between the minimum and the average, that is, it makes outliers much closer to the typical, average value, thus acting as a sort of regularizer.

A more stable measure is the standard deviation (Fig.~\ref{fig:std}). Forcing the potential to be more and more short-ranged (i.e. increasing $\alpha$) makes the peak sharper and sharper, decreasing the STD. A comparison of the values obtained shows that $V_3$ and $V_4$ perform better. This is reasonable, since these potentials cut interactions between far-away points, which we expect should matter little in the formation of a crystalline structure.

A clearer understanding is obtained considering the ratio STD/mean (Fig. \ref{fig:ratio}).
Of all the potentials taken into account, $V_3$ with $\alpha = 7$ is the one performing better. In general, for equal $\alpha$, $V_4$ performs better than $V_3$. However, due to the computational complications of dealing with exponentials, the former requires more care to be implemented at higher values of $\alpha$. All in all, it would seem that the $V_3$ is particularly well suited for the task at hand and it is, therefore, the one we studied more in depth.

\subsection{The $\alpha \to \infty$, $N_p \to \infty$ extrapolation}\label{sec:extrapolation}

\begin{figure}[t]
    \centering
    \includegraphics[width=0.75\textwidth]{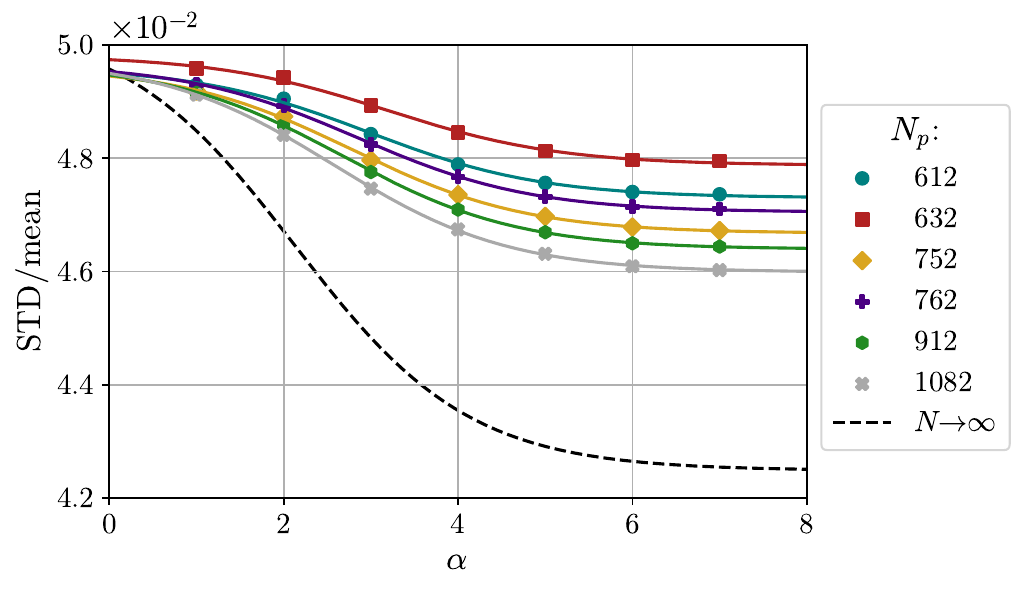}
    \caption[Ratio STD/mean as a function of $\alpha$ for different $N_p$]{Ratio STD/mean as a function of $\alpha$ for different $N_p$, together with sigmoidal fits and extrapolation at $N_p \to \infty$.}
    \label{fig:together}
\end{figure}

Since, in principle, we are interested in the limit of large $N_p$ and infinite $\alpha$ (sharp potential connected to Tammes problem), we want to quantify the limitations that arise from considering only finite values of $N_p$ and $\alpha$, and what is the difference from the asymptotic case.
First of all, we plot the ratio STD/mean as a function of $\alpha$ for different $N_p$, corresponding to different $m$ and $n$ (Fig. \ref{fig:together}). We considered only cases with $N_p$ large enough for the behavior to be monotonically decreasing and the cases with $m = n$ or at least similar since it has been conjectured that configurations with $m$ much greater than $n$ have high (Coulomb) energy and are far from optimal \cite{altschuler1997possible}.
We then performed a sigmoidal fit using a function $f(x; A, B, C) = C+ B[1 - S(x-A)]$, where $S = 1/(1+e^{-x})$. Extrapolation to large values of $\alpha$ gives variations of the order of 0.1 \% with respect to the $\alpha = 7$ case for all $N_p$. 

\begin{figure}[t]
    \centering
    \includegraphics[width=0.8\textwidth]{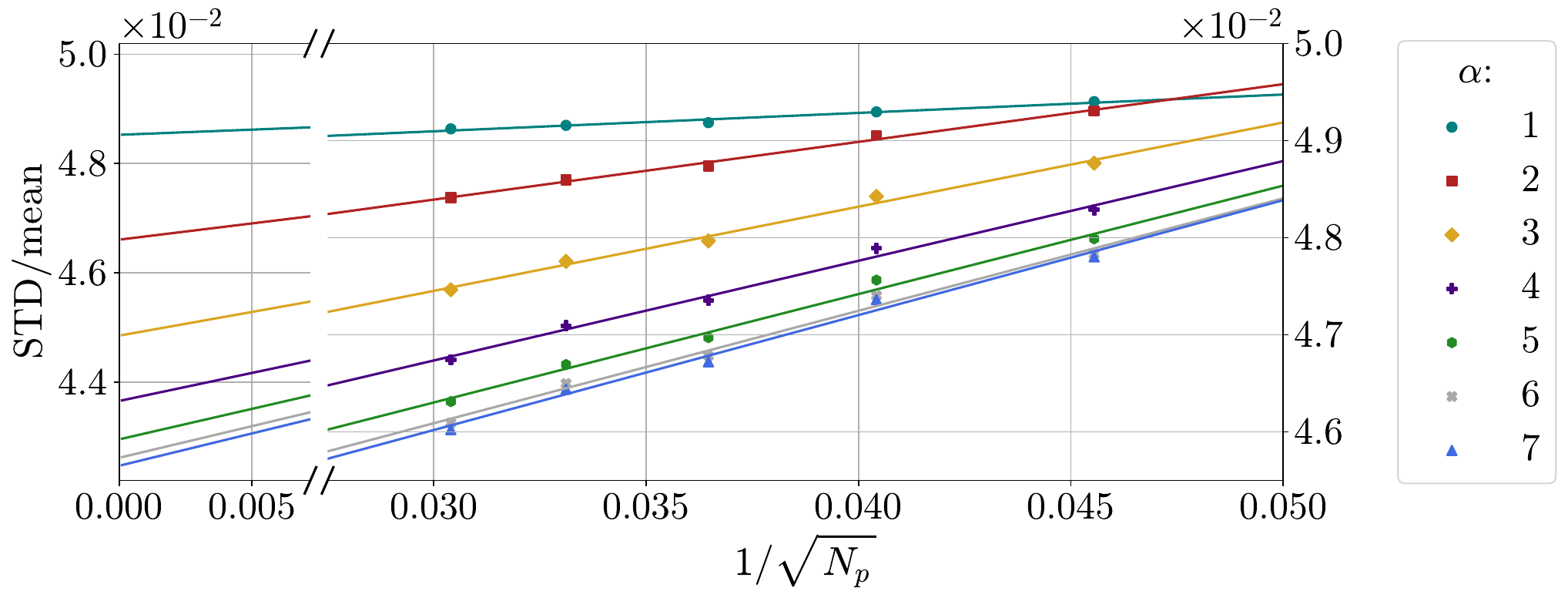}
    \caption[Ratio STD/mean as a function of $1/\sqrt{N_p}$ for different $\alpha$]{Ratio STD/mean as a function of $1/\sqrt{N_p}$ for different $\alpha$, together with fits and extrapolations at $N_p \to \infty$.}
    \label{fig:different_N}
\end{figure}

\begin{figure}[t]
    \centering
    \includegraphics[width=0.8\textwidth]{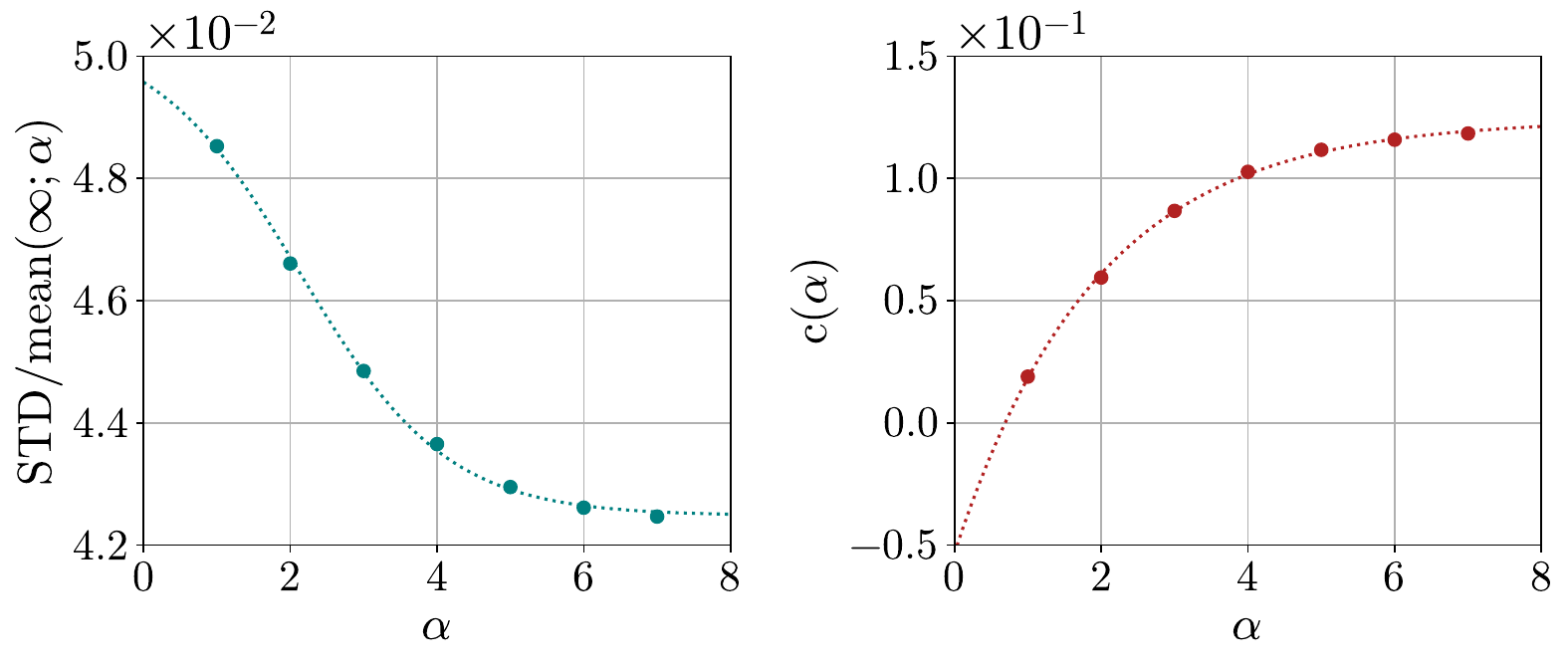}
    \caption[STD/mean($\infty; \alpha$) and $c(\alpha)$ coefficients as a function of $\alpha$]{STD/mean($\infty; \alpha$) and $c(\alpha)$ coefficients as a function of $\alpha$, together with sigmoidal and exponential fits (dotted lines).}
    \label{fig:asymptotics}
\end{figure}

For each $\alpha$ we then performed a fit using the function:
\begin{equation}
    \frac{\text{STD}}{\text{mean}}(N; \alpha) = \frac{\text{STD}}{\text{mean}}(\infty; \alpha) +\frac{c(\alpha)}{\sqrt{N_p}}.
\end{equation}
The results are presented in Fig. \ref{fig:different_N}.
We then considered the behaviour of $\frac{\text{STD}}{\text{mean}}(\infty; \alpha)$ and $c(\alpha)$ as functions of $\alpha$ (Fig. \ref{fig:asymptotics}) and performed sigmoidal (as before) and exponential fits, respectively.

Finally, in Fig. \ref{fig:together} $\frac{\text{STD}}{\text{mean}}(\infty; \alpha)$ is plotted together with the values obtained at finite $N_p$. 

In the end, the effect of finite $N_p$ is the one that seems to play the greatest role, with effects of the order of 10\%, while the effect of finite $\alpha$ is small. Thus, using values of $\alpha \geq 7$ should already yield good results, close enough to the infinitely sharp case.

\subsection{Stability}\label{sec:stability}

Another question that arises naturally when analyzing these distributions is whether they are stable under the application of a small perturbation. We studied the stability of the distributions obtained using the LP procedure and different values of $\alpha$. In order to do so, we used the following procedure:
\begin{enumerate}
    \item we generated $N_p$ points using the LP algorithm;
    \item to all points we applied a random Gaussian perturbation with zero mean and standard deviation equal to the 5\% of the mean distance between first neighbours, as obtained through \eqref{eq:char_len}. Then we projected the points back on the sphere;
    \item we performed GD using $V_3$ with different $\alpha$ and saw if the distribution of the standard deviation and of the ratio $\frac{\text{STD}}{\text{mean}}$ was the same as the one obtained without the addition of the perturbation.
\end{enumerate}

\begin{figure}[t]
     \centering
     \begin{subfigure}[b]{1 \textwidth}
    \centering
    \makebox[\textwidth][c]{\includegraphics[width=1.0\textwidth]{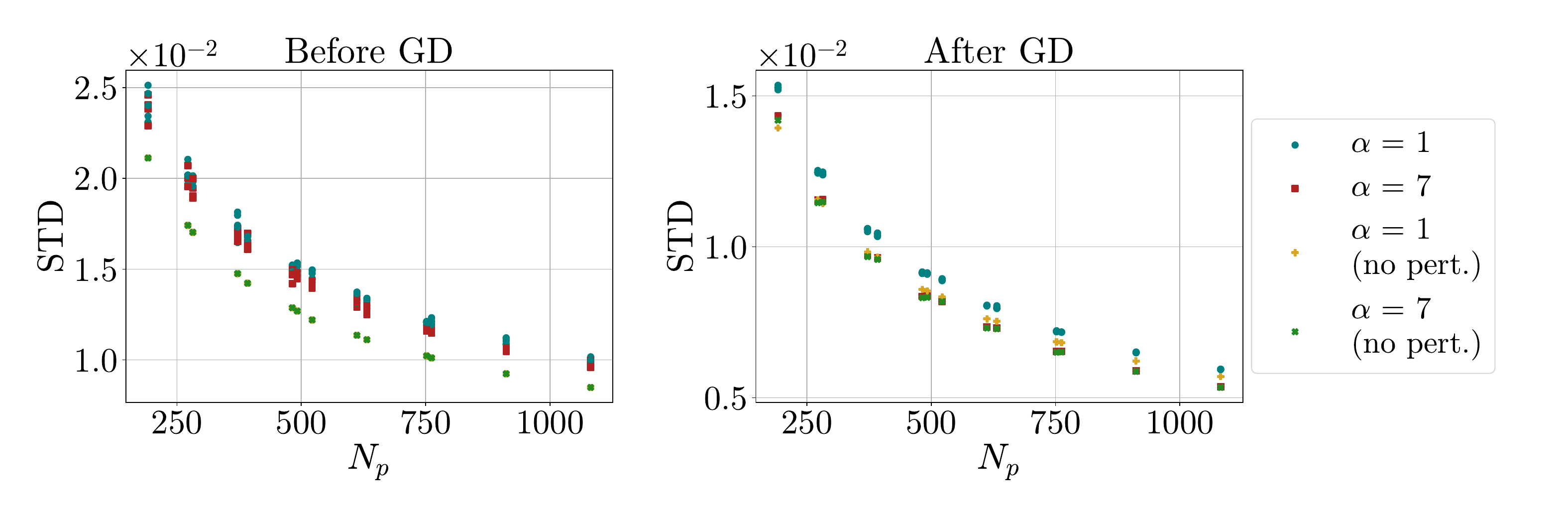}}%
    \caption{STD}
    \label{fig:primo}
     \end{subfigure}
     \hfill
     \begin{subfigure}[b]{1 \textwidth}
    \centering
    \makebox[\textwidth][c]{\includegraphics[width=1.0\textwidth]{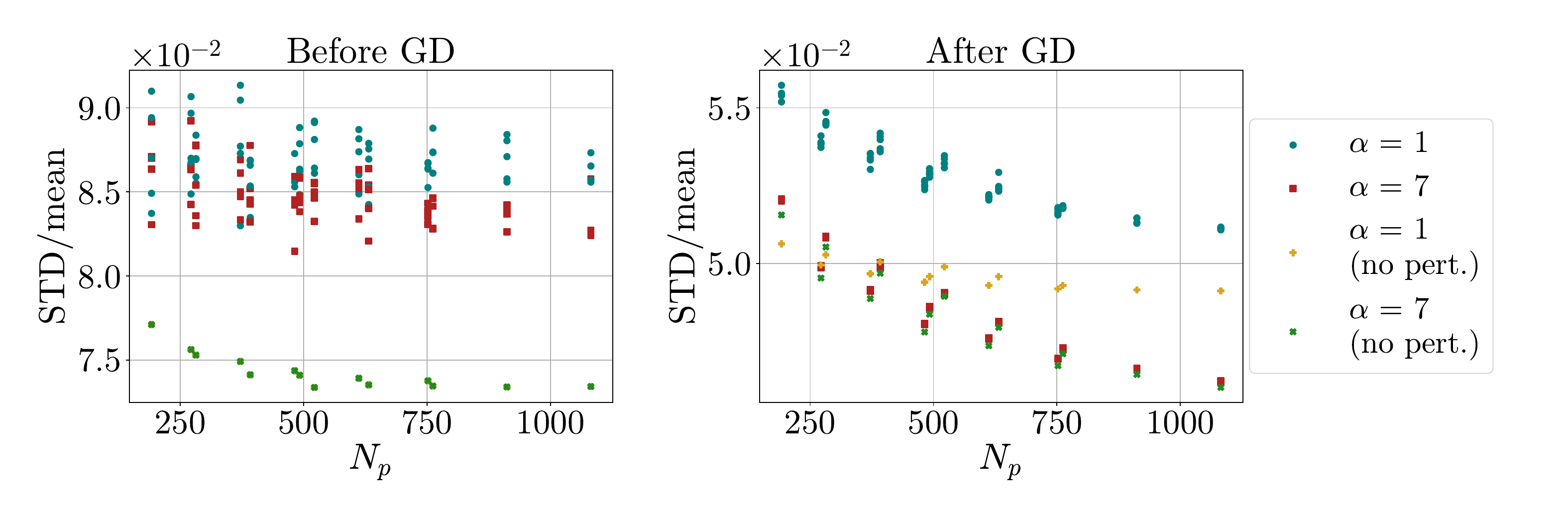}}%
    \caption{STD/mean}
    \label{fig:secondo}
     \end{subfigure}

        \caption[STD and STD/mean as a function of $N_p$ for distribution of points with added noise]{STD (a) and STD/mean (b) as a function of $N_p$ for distribution of points with added noise, before and after GD. Points (circles and squares) are perturbed configurations. Crosses are distributions of points after LP algorithm (left) and distributions after GD (right).}
        \label{fig:stability}

\end{figure}

We did these steps for $\alpha = 1, 7$, repeating the process 5 times for each value of $\alpha$.
Results are presented in Fig. \ref{fig:stability}.  The choice $\alpha = 7$ cancels differences much faster (all cases basically collapse on each other after GD) and recovers quickly the values obtained in absence of perturbation. We also performed the same procedure with a preliminary GD optimization between the first and the second steps. Results present no significant difference with respect to the shown case.

 All things considered, the $\alpha = 7$ case not only demonstrates a superior capability in reducing the ratio STD/mean, but it also appears to be considerably more stable.

\FloatBarrier

\section{Conclusions}\label{sec:conclusions}

In conclusion, in this work, we have introduced a framework to analyze the uniformity of distributions of point on the sphere. We introduce new measures alternative to those commonly found in literature. We believe the observables described in this paper are better descriptors of the proprieties of uniformity usually required in practical applications.

We have used these new tools to study a variety of algorithms that are aimed at producing uniform points on the sphere. Further we have optimized these point distributions through the minimization of several interaction potentials via the gradient descent algorithm.
Finally, we have characterized finite size effects in an extensive way.

In the end, we found that the Lattice Points algorithm followed by a GD optimization with power-law potential with a cutoff performs particularly well, and already the potential with $\alpha = 7$ seems to yield very good results. This is a very good compromise between the optimal potentials ($\alpha \gg 1$) and potentials smooth enough to be optimized quickly.

The methods used in this paper can be further generalized to study different algorithms and optimization methods, therefore they allow us to find the most uniform distribution of points on the sphere.

\section*{Acknowledgments}

We thank Rafael Diaz Hernandez Rojas for the fruitful discussion and support. The research has received financial support from ICSC - Italian Research Center on High-Performance Computing, Big Data, and Quantum Computing, funded by the European Union - NextGenerationEU.

\bibliographystyle{unsrt}
\bibliography{references}

\begin{thebibliography}{10}

\bibitem{thomson1904xxiv}
Joseph~John Thomson.
\newblock Xxiv. on the structure of the atom: an investigation of the stability
  and periods of oscillation of a number of corpuscles arranged at equal
  intervals around the circumference of a circle; with application of the
  results to the theory of atomic structure.
\newblock {\em The London, Edinburgh, and Dublin Philosophical Magazine and
  Journal of Science}, 7(39):237--265, 1904.

\bibitem{2b51ac61c9fb43a29f181b14458a1f5e}
{Pieter Merkus Lambertus} Tammes.
\newblock {\em On the origin of number and arrangement of the places of exit on
  the surface of pollen-grains}.
\newblock PhD thesis, 1930.
\newblock Relation: http://www.rug.nl/ Rights: De Bussy.

\bibitem{barreira2011surface}
Raquel Barreira, Charles~M Elliott, and Anotida Madzvamuse.
\newblock The surface finite element method for pattern formation on evolving
  biological surfaces.
\newblock {\em Journal of mathematical biology}, 63:1095--1119, 2011.

\bibitem{PhysRevLett.98.185502}
V.~L. Lorman and S.~B. Rochal.
\newblock Density-wave theory of the capsid structure of small icosahedral
  viruses.
\newblock {\em Phys. Rev. Lett.}, 98:185502, Apr 2007.

\bibitem{944811}
J.~Verhaevert, E.~Van~Lil, and A.~Van~de Capelle.
\newblock Uniform spherical distributions for adaptive array applications.
\newblock In {\em IEEE VTS 53rd Vehicular Technology Conference, Spring 2001.
  Proceedings (Cat. No.01CH37202)}, volume~1, pages 98--102 vol.1, 2001.

\bibitem{bauer2000distribution}
Robert Bauer.
\newblock Distribution of points on a sphere with application to star catalogs.
\newblock {\em Journal of Guidance, Control, and Dynamics}, 23(1):130--137,
  2000.

\bibitem{beentjes2015quadrature}
Casper~HL Beentjes.
\newblock Quadrature on a spherical surface.
\newblock {\em Working note available on the website http://people. maths. ox.
  ac. uk/beentjes/Essays}, 2015.

\bibitem{saff1997distributing}
Edward~B Saff and Amo~BJ Kuijlaars.
\newblock Distributing many points on a sphere.
\newblock {\em The mathematical intelligencer}, 19:5--11, 1997.

\bibitem{Y.Levin_2003}
Y.~Levin and J.~J. Arenzon.
\newblock Why charges go to the surface: A generalized thomson problem.
\newblock {\em Europhysics Letters}, 63(3):415, aug 2003.

\bibitem{PhysRevLett.89.185502}
M.~Bowick, A.~Cacciuto, D.~R. Nelson, and A.~Travesset.
\newblock Crystalline order on a sphere and the generalized thomson problem.
\newblock {\em Phys. Rev. Lett.}, 89:185502, Oct 2002.

\bibitem{altschuler1997possible}
Eric~Lewin Altschuler, Timothy~J Williams, Edward~R Ratner, Robert Tipton,
  Richard Stong, Farid Dowla, and Frederick Wooten.
\newblock Possible global minimum lattice configurations for thomson's problem
  of charges on a sphere.
\newblock {\em Physical Review Letters}, 78(14):2681, 1997.

\bibitem{erber1997complex}
T~Erber and GM~Hockney.
\newblock Complex systems: Equilibrium configurations of n equal charges on a
  sphere (2 ≤ n ≥ 112).
\newblock {\em Advances in chemical physics}, 98:495--594, 1997.

\bibitem{simAnnealing}
Yang Xiang, Sylvain Gubian, Brian Suomela, and Julia Hoeng.
\newblock Generalized simulated annealing for global optimization: The gensa
  package.
\newblock {\em The R Journal Volume 5(1):13-29, June 2013}, 5, 06 2013.

\bibitem{Gorski_2005}
K.~M. Gorski, E.~Hivon, A.~J. Banday, B.~D. Wandelt, F.~K. Hansen, M.~Reinecke,
  and M.~Bartelmann.
\newblock {HEALPix}: A framework for high-resolution discretization and fast
  analysis of data distributed on the sphere.
\newblock {\em The Astrophysical Journal}, 622(2):759--771, apr 2005.

\bibitem{malkin2019new}
Zinovy Malkin.
\newblock A new equal-area isolatitudinal grid on a spherical surface.
\newblock {\em The Astronomical Journal}, 158(4):158, 2019.

\bibitem{del2024compute}
Luca~Maria Del~Bono, Flavio Nicoletti, and Federico Ricci-Tersenghi.
\newblock How to compute efficiently the analytical solution to heisenberg spin
  glass models on sparse random graphs and their de almeida-thouless line.
\newblock {\em arXiv preprint arXiv:2406.16836}, 2024.

\bibitem{bovzivc2019spherical}
An{\v{z}}e~Lo{\v{s}}dorfer Bo{\v{z}}i{\v{c}} and Simon {\v{C}}opar.
\newblock Spherical structure factor and classification of hyperuniformity on
  the sphere.
\newblock {\em Physical Review E}, 99(3):032601, 2019.

\bibitem{meyra2019hyperuniformity}
Ariel~G Meyra, Guillermo~J Zarragoicoechea, Alberto~L Maltz, Enrique Lomba, and
  Salvatore Torquato.
\newblock Hyperuniformity on spherical surfaces.
\newblock {\em Physical Review E}, 100(2):022107, 2019.

\bibitem{chang2010simple}
Hai-Chau Chang and Lih-Chung Wang.
\newblock A simple proof of thue's theorem on circle packing.
\newblock {\em arXiv preprint arXiv:1009.4322}, 2010.

\bibitem{wales2006structure}
David~J Wales and Sidika Ulker.
\newblock Structure and dynamics of spherical crystals characterized for the
  thomson problem.
\newblock {\em Physical Review B}, 74(21):212101, 2006.

\bibitem{perez1997influence}
Antonio P{\'e}rez-Garrido, MJW Dodgson, and MA~Moore.
\newblock Influence of dislocations in thomson’s problem.
\newblock {\em Physical Review B}, 56(7):3640, 1997.

\bibitem{katanforoush2003distributing}
Ali Katanforoush and Mehrdad Shahshahani.
\newblock Distributing points on the sphere, i.
\newblock {\em Experimental Mathematics}, 12(2):199--209, 2003.

\bibitem{perez1999symmetric}
A~P{\'e}rez-Garrido and MA~Moore.
\newblock Symmetric patterns of dislocations in thomson’s problem.
\newblock {\em Physical Review B}, 60(23):15628, 1999.

\bibitem{wales2009defect}
David~J Wales, Hayley McKay, and Eric~L Altschuler.
\newblock Defect motifs for spherical topologies.
\newblock {\em Physical Review B}, 79(22):224115, 2009.

\bibitem{altschuler2006defect}
Eric~Lewin Altschuler and Antonio P{\'e}rez-Garrido.
\newblock Defect-free global minima in thomson’s problem of charges on a
  sphere.
\newblock {\em Physical Review E}, 73(3):036108, 2006.

\bibitem{alishahi2015spherical}
Kasra Alishahi and Mohammadsadegh Zamani.
\newblock The spherical ensemble and uniform distribution of points on the
  sphere.
\newblock 2015.

\bibitem{Ashcroft76}
N.~W. Ashcroft and N.~D. Mermin.
\newblock {\em {S}olid {S}tate {P}hysics}.
\newblock Holt-Saunders, 1976.

\bibitem{reference.wolfram_2022_spherepoints}
Wolfram Research.
\newblock {SpherePoints}.
\newblock \url{https://reference.wolfram.com/language/ref/SpherePoints.html},
  2022.

\bibitem{Tegmark_1996}
Max Tegmark.
\newblock An icosahedron-based method for pixelizing the celestial sphere.
\newblock {\em The Astrophysical Journal}, 470(2):L81, oct 1996.

\end{thebibliography}

\end{document}